\documentclass{article}
\usepackage{arxiv}
\usepackage{graphicx,amsmath,amsfonts,amscd,amsthm,amssymb,pstricks}
\usepackage{color}
\usepackage{dcolumn}
\usepackage{bm}
\usepackage{longtable}
\usepackage{mathrsfs}
\usepackage{textcomp}
\usepackage{esvect}
\usepackage{float}
\usepackage[USenglish,british]{babel}
\usepackage{environ}
\usepackage{xifthen}
\usepackage{xargs}		
\usepackage{hyperref}
\usepackage[separate-uncertainty = true]{siunitx}
\usepackage{physics,mathtools}
\usepackage{multirow}
\usepackage{comment}
\usepackage{enumerate}
\usepackage{appendix}
\usepackage{color}
\usepackage{booktabs}
\usepackage{tikz}
\usetikzlibrary{arrows, calc, matrix, patterns, decorations.markings, shapes, decorations.pathmorphing, shadows.blur}
\usepackage[utf8]{inputenc}
\usepackage{enumitem}
\usepackage[separate-uncertainty = true]{siunitx}

\DeclareMathOperator{\arcsinh}{arcsinh}

\usepackage{xparse}

\title{Harmonic analysis of non-stationary signals with application to LHC beam measurements}

\author{G. Russo\\
Beams Department, CERN, Esplanade\ des Particules 1, 1211 Geneva 23, Switzerland\\
Goethe University, 60323 Frankfurt am Main, Germany\\
G. Franchetti\\
GSI Helmholtzzentrum f\"ur Schwerionenforschung GmbH, Planckstraße 1, 64291 Darmstadt, Germany\\
M. Giovannozzi\thanks{Corresponding author: massimo.giovannozzi@cern.ch}\\
Beams Department, CERN, Esplanade\ des Particules 1, 1211 Geneva 23, Switzerland\\
E. H. Maclean\\
Beams Department, CERN, Esplanade\ des Particules 1, 1211 Geneva 23, Switzerland}

\begin{document}
\maketitle

\begin{abstract}
Harmonic analysis has provided powerful tools to accurately determine the tune from turn-by-turn data originating from numerical simulations or beam measurements in circular accelerators and storage rings. Methods that have been developed since the 1990s are suitable for stationary signals, i.e., time series whose properties do not vary with time and are represented by stationary signals. However, it is common experience that accelerator physics is a rich source of time series in which the signal amplitude varies over time. Furthermore, the properties of the amplitude variation of the signal often contain essential information about the phenomena under consideration. In this paper, a novel approach is presented, suitable for determining the tune of a non-stationary signal, {which is based on the use of the Hilbert transform}. The accuracy of the proposed methods is assessed in detail, and an application to the analysis of beam data collected at the CERN Large Hadron Collider is presented and discussed in detail.
\end{abstract}


\section{introduction} \label{sec:intro}

It is well known that the dynamics of a charged particle under the influence of the nonlinear fields generated by the magnetic lattice of a circular accelerator can be described in the form of a quasi-periodic function
\begin{equation}
\begin{split}
    z(n) = & u(n) -i p_u(n) = \\
    = & \sum_{j=1}^M a_j \exp{2 \pi i \omega_j n} \, , \,\, a_j \in \mathbb{C}, \,\, \omega_j\in \mathbb{R} \, , 
\end{split}
\label{eq:stationary}
\end{equation}
where {$n$ is the number of turns,} $u, p_u$ represent conjugate variables ($u$ stands for $x$ or $y$), and $\omega_j, \, \, 1 \leq j \leq M$ are the frequencies describing the particle's motion, and $a_j$ the corresponding amplitudes. {The tune is represented by the $\omega_j$ that corresponds to the largest $\vert a_j \vert$. It is customary to order the frequencies by decreasing values of $\vert a_j \vert$ so that the tune corresponds to $\omega_1$.} The properties of the orbit are closely related to the {set of}  $\omega_j$, and their determination from knowledge of the time series $z(n)$ is a fundamental problem that is addressed by harmonic analysis. 

{In the general case of 4D betatron oscillations, the motion is described by two vectors 
\begin{equation}
    \begin{split}
        z(n) & =x(n)-i p_x(n) \\
        w(n) & =y(n)-i p_y(n) \, , 
    \end{split}
\end{equation}
each vector $z(n)$ and $w(n)$ being represented in the form given in Eq.~\eqref{eq:stationary}. The tune in each plane is given by the frequency with the highest amplitude within that plane. Other frequency components may arise due to various types of coupling, whether linear or nonlinear, between the $x$ and $y$ planes. However, as long as these components are well separated from the tunes and possess lower amplitudes, the coupling effects can be ignored, making the one-dimensional analysis presented in the remainder of this paper valid.}

The fundamental tool for harmonic analysis is the fast Fourier transform ({DFT}) that provides the frequency spectrum of $z(n)$. However, each frequency component of the Fourier spectrum is determined with an accuracy that scales as $1/N$, where $N$ represents the maximum number of points in the time series $z(n)$. 

In the 1990s, intense efforts were devoted to improve methods for computing $\omega_j$ with a precision higher than that provided by {DFT}, i.e., $1/N$. In several cases, these developments originated in the field of celestial mechanics~\cite{laskar:1992,laskar:1993,Laskar:1999,Papaphilippou:1999,Laskar:2003a} and were then promoted to the field of accelerator physics (see, e.g., Ref.~\cite{Laskar:2019}). These algorithms have been implemented in a number of codes that are now of standard use for the analysis of beam dynamics~\cite{Bartolini:1997np,Bartolini:702438} and in the analysis of massive simulation data in the form of what is called frequency map analysis (FMA)~\cite{papaphilippou:1996,papaphiluppou:1998,laskar:epac00-tup3a15,turchetti:2001a,turchetti:2001b,belgroune:epac02-weple011,laskar:pac03-woab001,papaphilippou:pac03-rppg007,laskar:2003b,papaphilippou:epac04-weplt084,papaphilippou:2014}. 

Independently, alternative techniques were developed, based on the closed-form formula approach that provides the value of $\omega_j$ as a refinement of the estimate obtained from the application of the {DFT} algorithm~\cite{Bartolini:292773}, and this is the approach that will be followed in the rest of this article. {\color{black} It is important to note that this method has not been extensively explored, and we leverage these recent studies to advance it further.} The key results can be summarized as follows: The tune estimate provided by the {DFT} spectrum can be improved to achieve an accuracy that scales as $1/N^2$ (this approach will be called interpolated {DFT} in the rest of the paper). Moreover, when the Hanning filter is applied to the original time series, the tune estimate provided by the {DFT} spectrum can be improved to an accuracy that scales as $1/N^4$ (this approach will be called interpolated {DFT} with the Hanning filter in the rest of the paper). 

It is also worth mentioning that methods based on {DFT} are not the only options available to compute $\omega_j$ with high accuracy. In fact, the average phase advance~\cite{Bartolini:292773} is an excellent alternative that has been further developed in recent years~\cite{russo:ipac21-thpab189}. However, this approach will not be discussed or used in this paper. {The primary justification for this selection is that DFT-based methods allow for the analysis of the entire spectrum of the time series, unlike the average phase advance, which yields only the dominant frequency, namely, the tune.}

When other phenomena are considered, e.g., instead of a single particle an ensemble of particles is considered to mimic the result of a beam measurement, or a coupling between transverse and longitudinal dynamics is considered, {the measured turn-by-turn signal features decoherence and} the model of Eq.~\eqref{eq:stationary} is no longer valid. An appropriate description is then given, in general terms, by
\begin{equation}
\begin{split}
    z(n) = & u(n) -i p_u(n) = f(n) \times \\
    & \times \left (\sum_{j=1}^M a_j \exp{2\pi i \omega_j n} \right )\,\, a_j \in \mathbb{C}, \,\, \omega_j\in \mathbb{R} \, . 
\end{split}
\label{eq:nonstationary}
\end{equation}

In this case, the signal $z(n)$ is non-stationary and the function $f(n)$ represents a modulation of the signal amplitude. It is important to stress that, in general, $f(n)$ contains key information about the physical process that governs the dynamics, and determining its form and the parameters that describe it is essential. Furthermore, applying the {DFT} approach (or any {DFT}-based approach) does not provide the correct answer to the problem of determining $\omega_j$. In this paper, we show how it is possible to devise an appropriate technique that is capable of determining not only $\omega_j$, but also the physical parameters that model $f(n)$.

The structure of the paper is as follows: In Section~\ref{sec:models},  phenomena that generate non-stationary signals are presented and discussed in detail, with particular emphasis on the dependence of the amplitude variation of the key theoretical parameters describing the phenomenon under consideration. In Section~\ref{AnalyticalEquationsForDampedExponential} the novel method is presented and discussed in detail.  The comparison of the accuracy of the various methods is presented in Section~\ref{NumericalSimulations}. In Section~\ref{DecoherenceProtonBeams_Section} the novel techniques are used for the analysis of a set of beam measurements performed at the CERN Large Hadron Collider (LHC) in 2012~\cite{EwenPaper} that show very promising reconstruction performance. Finally, conclusions are presented in Section~\ref{sec:conc}, and mathematical details are provided in the appendices. 

\section{Models of kicked-beam dynamics and tune determination}\label{sec:models}

In a circular accelerator, the transverse tune is measured by kicking the beam in a given transverse plane, so that the subsequent oscillations are recorded on a turn-by-turn basis. Therefore, analysis of the accuracy of tune determination in a storage ring or a collider requires modeling the evolution of the transverse beam barycenter of kicked beams. In general, the measured signal represents the barycenter of the charge distribution measured by a beam position monitor (BPM), and it features amplitude variations. There are several phenomena that can generate such oscillations, and in the following some models will be described from a mathematical and accelerator physics point of view to highlight the essential parameters and define their typical values for real-life applications.

\begin{figure}
    \centering
    \includegraphics[trim=40mm 31mm 52mm 43mm,width=\columnwidth,clip=]{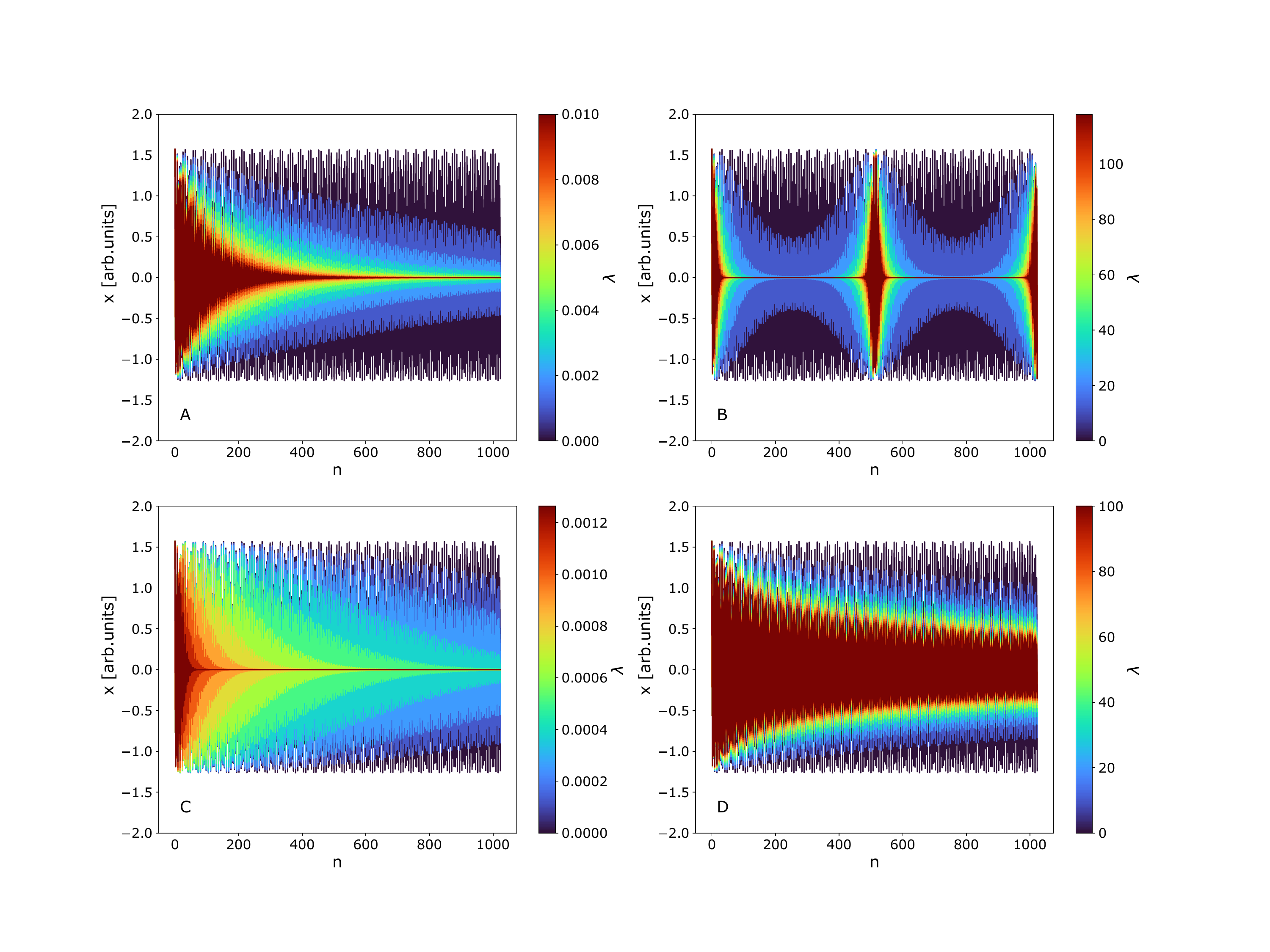}
    \caption{Simulated signals representing the oscillation of the barycenter of a kicked beam as a result of energy damping (A), chromatic decoherence (B), nonlinearities (C), and acceleration (D). These signals are shown as functions of the number of turns $n$ (up to the maximum of $1024$ turns). For each signal, the characteristic scaling factor $\lambda$ has been varied in the interval shown on the color scale.} 
    \label{fig:RealSignals}
\end{figure}

Figure~\ref{fig:RealSignals} shows the turn-by-turn evolution of the barycenter of the beam for the four damping phenomena analyzed in the following. Note that the general model of the oscillation of a kicked beam is given in Eq.~\ref{eq:nonstationary}, 
however, as we will discuss in the following, the function $f(n)$ will take an exponential form 
\begin{equation}
   {f(n) \propto e^{-\lambda g(n)} \, .} 
\end{equation}

In the four panels, the time dependence of the oscillations is shown, together with the dependence of the damping characteristics on the key parameter $\lambda$, which varies for the four phenomena under consideration. 

\subsection{Amplitude variation from energy damping} \label{ExponentialDamp}
An exponential damping of the signal amplitude is typical of kicked electron beams, where, under the assumption of linear transverse beam dynamics, which implies that no amplitude detuning is present, the transverse amplitude decays exponentially due to synchrotron radiation. The possibility of determining the tune and damping factor using an analytical approach makes this case very interesting. The turn-by-turn evolution of the beam barycenter can be described as:
\begin{equation}\label{ElectronBeamEvolution}
    x(n) = \sqrt{2 J_x(n) \beta_x} \cos 2 \pi n \nu \,,
\end{equation}
where $2 J_x(n)$, describes the turn-by-turn evolution of the single-particle emittance of an electron beam, defined as:
\begin{equation} \label{ActionEvolutionElectronBeam}
    J_x(n) = J_x(0) \exp \left( - \lambda n \right)\, ,
\end{equation}
where generic damping parameter $\lambda$ becomes
\begin{equation}
    \lambda = \frac{j_x U_0}{2 E_0 T_0} \, ,
\end{equation}
with $T_0$ the revolution period, $U_0$ the energy loss per unit turn, $E_0$ the reference energy of the particle, and 
\begin{equation}
{j_x = 1 - \frac{I_4}{I_2}}
\end{equation}
is the damping partition number defined in terms of the second ($I_2$) and fourth ($I_4$) synchrotron radiation integral~\cite{Book_AcceleratorPhysics}. By performing a re-normalization of Eq.~\eqref{ElectronBeamEvolution}, the beam barycenter reads as:
\begin{equation}
    \Tilde{x}(n) = \frac{x(n)}{\sqrt{2\beta_x J_x(n=0)}} =\exp \left( - \lambda/2 n \right) \cos 2 \pi n \nu \, .
    \label{eq:energy-damping}
\end{equation}

In an ideal case where there is no damping, $\lambda$ is zero, otherwise $\lambda$ can range {in the interval
\begin{equation}
    1 \times 10^{-5} \leq \lambda \leq 1 \times 10^{-2} \, , 
\end{equation}
the lower value corresponding to the case of the CERN Electron Positron Accumulator (EPA)~\cite{Delahaye:2840520} and the upper value to the CERN Large Electron Positron (LEP) ring~\cite{LepDesignReport}. The given range is broad enough to encompass scenarios related to various accelerators globally.}

Figure~\ref{fig:RealSignals}~(A) represents the evolution of the BPM signal according to the model~\eqref{eq:energy-damping}, in which $\lambda$ is varied from zero, corresponding to a constant-amplitude signal, to the highly-damped signal, corresponding to the LEP case. 


\subsection{Amplitude variation from chromatic decoherence}\label{ChromaticDecoherence}

The model describing the turn-by-turn evolution of a kicked beam under the influence of chromatic decoherence was first described by Meller et al. in Ref.~\cite{Meller:1987ug}, where the evolution of the centroid of a proton beam after a transverse kick is described considering the effect of energy spread and transverse tune variation due to linear chromaticity. 

The energy spread does not cause a change in the betatron tune of the centroid, but rather creates a periodic amplitude modulation of the beam centroid that reads as:
\begin{equation}\label{ChromaticBeamCentre}
    \Tilde{x}(n) = \exp \left ( - \lambda \sin^2 \pi\nu_{{\mathrm{s}}} n \right) \cos 2\pi n \nu \, ,
\end{equation}
where the generic damping parameter $\lambda$ becomes
\begin{equation}\label{LambdaChromaticDecoherence}
    \lambda = \frac{1}{2} \left( 2 \frac{\sigma_\delta \xi}{\nu_{\mathrm{s}}} \right)^2\, ,
\end{equation}
and $\sigma_\delta$, $\xi=dQ/d \delta$ and $\nu_{\mathrm{s}}$ are the rms {relative momentum spread}, the horizontal chromaticity, $\delta$ is the relative momentum offset, and the synchrotron tune, respectively. {Note that $\nu_\mathrm{s}$ is expressed in units of turns and is hence dimensionless.}

These parameters vary a lot between accelerators and even for the same ring, depending on the type of beam produced. 

Figure~\ref{fig:RealSignals}~(B) represents the evolution of the BPM signal according to the model~\eqref{ChromaticBeamCentre} using typical beam parameters from the CERN Proton Synchrotron (PS) to determine the value of $\lambda$. It should be noted the peculiar behavior of the BPM signal, with periodic increases and decreases in the envelope amplitude generated by the periodic term in the argument of the exponential in Eq.~\eqref{ChromaticBeamCentre}. 

\subsection{Amplitude variation from nonlinearities}\label{NonLinearDecoherence}

There are several sources of nonlinearities in an accelerator ring, but here we refer to nonlinear magnetic-field errors. Under the influence of these errors, when a proton beam is kicked at a given amplitude, its transverse distribution will undergo filamentation, thus occupying a larger region in phase space~\cite{Book_AcceleratorPhysics}. The decoherence of betatron oscillations under the influence of nonlinear effects was first described by Meller et al. in Ref.~\cite{Meller:1987ug}, where a Gaussian transverse beam distribution was assumed. Note that more refined models have been derived and discussed in Refs.~\cite{lee:1991,sargsyan:2011}. Its turn-by-turn evolution after a transverse kick can be analytically determined under the assumption that the tune depends on the action {$J$, which is expressed in units of the beam size,} as
\begin{equation}
    \nu(J)= \nu + \mu J \, , 
    \label{eq:amplitude-detuning}
\end{equation}
and where $\mu$ represents the amplitude detuning {and is dimensionless}. The beam centroid evolution can be described as follows
\begin{equation}\label{NonLinearDecoherenceBeamCentroid}
    x(n) = \sigma_x Z A(n) \cos \left( 2 \pi n \nu + \Delta\phi(n) \right)\, ,
\end{equation}
where $\Delta\phi(n)$ is the phase-shift of the centroid, $Z$ the kick amplitude {expressed in units of beam size, which makes $Z$ dimensionless}, and $\sigma_x$ the transverse rms beam size. $A(n)$ is the decoherence factor, in a form that depends on the amplitude of the kick and, if it is larger than the beam size, which is the case in many applications, one obtains the following
\begin{equation} \label{AmplitudeModulationAmplDetuning}
    A(n) = \text{exp} \left [ -\frac{1}{2} \left( 4\pi\mu Z n \right)^2 \right ]\, .
\end{equation}

By normalizing Eq.~\eqref{NonLinearDecoherenceBeamCentroid} with respect to the beam size and the kick, the evolution of the beam centroid reads as
\begin{equation}
    \Tilde{x}(n) = \frac{x(n)}{\sigma_x Z} = \text{exp} \left( - \lambda  n ^2 \right) \cos \left ( 2 \pi n \nu  + \Delta\phi(n)\right )\, ,
\end{equation}
where generic damping parameter $\lambda$ becomes
\begin{equation}
    \lambda = 8 \pi^2 \mu^2 Z^2 \, ,
    \label{eq:lambda-model-C}
\end{equation}
where one notes that $\lambda$ does not depend on the sign of $\mu$.

The decoherence of betatron oscillations is characterized by a Gaussian amplitude modulation due to the accelerator nonlinearities. The range of values of $\mu$ used in the numerical simulations has been taken from those measured in the PS and LHC rings at their injection energies, namely up to a maximum of 
\begin{equation}
    {\mu = 10^{-3}} \, .
\end{equation} 

In Fig.~\ref{fig:RealSignals}~(C), a typical example of signal decoherence is shown due to the presence of amplitude detuning, in which the fast Gaussian decay is clearly visible. 

\subsection{Amplitude variation from acceleration} \label{BeamCentroidDuringAcceleration}

The physical beam emittance is not a constant of motion during acceleration. In this case, the Lorentz factor $\gamma_{\mathrm{rel}}$, the relative speed $\beta_{\mathrm{rel}}$, and the normalized beam emittance $\epsilon^*$ fully describe the emittance damping phenomenon as
\begin{equation}
    \epsilon_x = \frac{\epsilon^*_{x}}{\beta_{\mathrm{rel}} \gamma_{\mathrm{rel}}}\, ,
\end{equation}
from which one obtains that the amplitude of the coherent oscillations of the kicked beam varies as
\begin{equation}
    x(n) = x(0) \sqrt{\frac{\beta_{\mathrm{i, rel}} \gamma_{\mathrm{i, rel}}}{\beta_{\mathrm{rel}}(n) \gamma_{\mathrm{rel}}(n)}} \cos 2\pi n \nu \, ,
\end{equation}
where the subscript $\mathrm{i}$ refers to the initial value of the Lorentz factor and relative speed when the beam was displaced to an amplitude $x(0)$, while $\beta_{\mathrm{rel}}(n), \gamma_{\mathrm{rel}}(n)$ refer to the turn-by-turn evolution of the same quantities. These parameters may vary in several different ways depending on the form of the energy acceleration used in the ring. If a linear variation of the beam momentum is performed from $p_{\mathrm{i}}$ to $p_{\mathrm{f}}$ over $N$ turns, the turn-by-turn damping of the coherent betatron oscillations reads
\begin{equation}
\begin{split}
    \Tilde{x}(n) & =  \frac{1}{\displaystyle{\sqrt{1+\frac{1-n}{1-N} \left ( \frac{p_\mathrm{f}}{p_\mathrm{i}}-1 \right )}}} \cos 2\pi n \nu \\
    & = \frac{1}{\displaystyle{\sqrt{1+\frac{1-n}{1-N} \lambda}}} \cos 2\pi n \nu \\
    & = \exp \left [-\frac{1}{2} {\log{ \left ( 1+\frac{1-n}{1-N} \lambda \right )}} \right ] \cos 2\pi n \nu \, ,
\end{split}
\end{equation}
where $\lambda$ is a scaling factor related to the Lorentz factor and $\beta_\mathrm{rel}$. It should be noted that among the mechanisms of damped transverse oscillations, this is the slowest, with a logarithmic dependence on the turn number $n$. 

In Fig.~\ref{fig:RealSignals}~(D), an example of damped oscillations during beam acceleration is presented. The slow decay, slower than that of the other mechanism, is clearly visible.

\subsection{Assessment of accuracy of the tune determination}

The damping phenomena described in the previous sections have a characteristic scaling factor $\lambda$ and a different dependence on the number of turns $n$ that will generally be called from now on $f(\lambda, n)$. Therefore, it is possible to write a general analytical expression that combines all the damping phenomena considered, as
\begin{equation}\label{SignalTune_OnlyRealPart}
    x(n) =
  a e^{-f(\lambda, n)} \left( \cos 2\pi n \nu  + \sum_{k=1}^4 a_k \cos 2\pi n k \nu \right ) \, ,
\end{equation}
where 
\begin{equation}
    \begin{split}
            {a_k} & {= e^{-k}} \\
            {\nu} & {= 6.281} \, ,
    \end{split}
\end{equation}
{where the value of $\nu$ is representative for the CERN PS ring.} 
%
{Four additional} harmonics have been added, representing additional terms of the measured signal (see, e.g.,~\cite{Bartolini:292773}). 

\begin{figure}
    \centering
    \includegraphics[trim=65mm 2mm 68mm 15mm,width=\columnwidth,clip=]{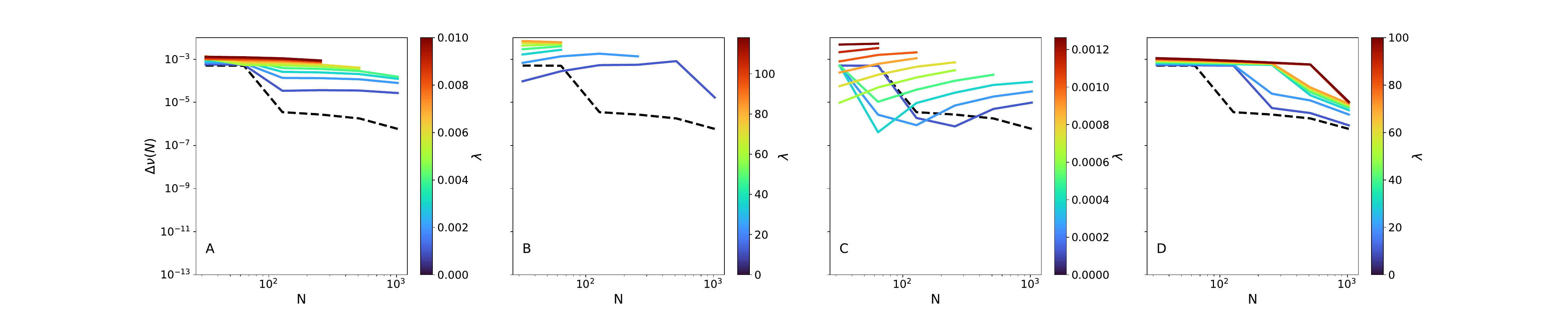}
    \includegraphics[trim=65mm 2mm 68mm 15mm,width=\columnwidth,clip=]{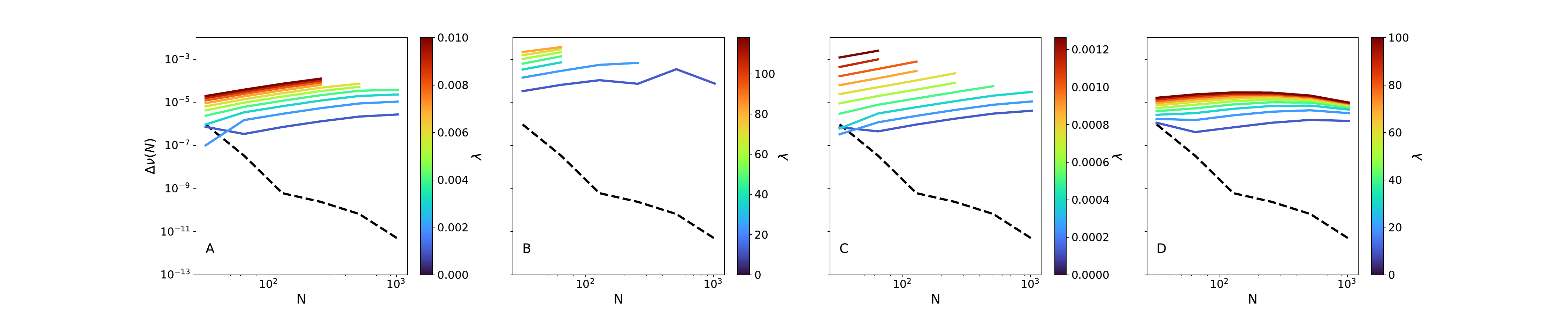}
    \caption{The tune error $\Delta \nu(N)$ is computed (from left to right) for the signals with amplitude variation from energy damping (A), chromatic decoherence (B), nonlinearities (C), and acceleration (D). Two methods have been used, namely, the interpolated {DFT} (top) and the interpolated {DFT} with Hanning filter (bottom). {The scaling of the tune error corresponding to the case $\lambda=0$ is represented by the back dashed lines with a dependence as $1/N^2$ (top) and $1/N^4$ (bottom). When $\lambda \neq 0$ the colored curves representing the tune error show a very different scaling with respect to the case of stationary signals.}} 
    \label{fig:TuneErrorConstantAmplitudeMethods}
\end{figure}

The signals generated by the four damping mechanisms have been analyzed with the methods briefly reviewed in Section~\ref{sec:intro} to determine the impact of modulation of the amplitude on the reconstruction of the tune. Note that accuracy is defined as the error in the calculation of the tune, namely 
\begin{equation}
    {\Delta \nu(N)= \nu(N) -\nu \, ,}
\end{equation}
where $\nu$ is the value of the tune used to generate the signal and $\nu(N)$ is the tune obtained using the reconstruction methods applied to a signal of length $N$. The results of this analysis are reported in Fig.~\ref{fig:TuneErrorConstantAmplitudeMethods} for two cases: interpolated {DFT} (top) and interpolated {DFT} with the Hanning filter (bottom), where it is clearly visible that the limit $\lambda \to 0$ is often discontinuous. {For comparative purposes, the dashed lines depict the curves $1/N^2$ (top) and $1/N^4$ (bottom), which illustrate the scaling of the tune error for the benchmark methods, specifically the Interpolated DFT and the Interpolated DFT with Hanning filter, respectively.}

The performance of these two methods is very different and critically depends on the damping mechanism and the value of the damping parameter. For the sake of comparison, the performance for the case without amplitude variation is also shown. In general, one observes that the error hardly decreases as a function of $N$ when the damping mechanism is different from zero. Furthermore, $\Delta \nu(N)$ tends to increase with $N$ for increasing damping effects. These observations clearly indicate that the application of the {interpolated DFT and interpolated DFT with Hanning filter}, derived for constant-amplitude signals, to varying-amplitude signals is not justified and generates errors in the tune reconstruction. In turn, these observations support the need for alternative approaches to deal with varying-amplitude signals, which will be presented and discussed in the next section. 
%
\section{Novel methods for determining the tune of a varying-amplitude signal} \label{AnalyticalEquationsForDampedExponential}

In this section, new closed-form solutions are presented to determine $\nu$ and $\lambda$ for a signal with exponential damping. Furthermore, the general method based on the use of the Hilbert transform~\cite{Book_Hilbert,luo:2009} for amplitude-varying signals is discussed.

\subsection{Closed-form solution for tune determination of exponentially damped signals} \label{sec:closed-form-damped}

Considering a signal of the form
\begin{equation}\label{Definition_DampExp}
    x(n) =
  a e^{-\lambda n} \cos 2\pi n \nu \, ,
\end{equation}
it is possible to obtain a closed-form solution to compute the value of $\nu$ and $\lambda$ using an interpolated {DFT} or an interpolated {DFT} with Hanning filter approaches and in doing so, we take advantage of previous studies carried out in 2015 and discussed in Ref.~\cite{Fabre:2043807}. 

For the signal defined in Eq.~\eqref{Definition_DampExp}, the {DFT} coefficients read
\begin{equation}\label{DampedExp_phi_definition}
  \phi (\nu _j) = \frac{1}{N} \sum_{n=1}^{\text{N}}
  e ^{2 \pi i (\nu - \nu _j) n - \lambda n} \, ,
\end{equation}
from which one obtains
\begin{equation}\label{Phi_Simplified_DampedExp}
  \left| \phi(\nu_j) \right|^2 =
   \frac{1}{N^2} e^{-\lambda(N+1)}
   \frac{\sin^2 \left( \pi N \Delta \nu_j \right) +
     \sinh^2 \frac{\lambda N }{2}   }
   {\sin^2 \left( \pi \Delta \nu_j  \right) +
     \sinh^2 \frac{\lambda }{2}} \, ,
\end{equation}
where 
\begin{equation}
    {\Delta \nu_j = \nu - \nu _j \, .}
\end{equation}

If $k$ indicates the index at which $\left| \phi(\nu_j) \right|^2$ reaches its maximum, it is possible to use the Fourier components corresponding to the indices $k, k \pm 1$ to determine $\nu$ and $\lambda$, which are given in terms of two solutions, namely
\begin{equation}\label{Tune_DampedExponential}
    \begin{split}
       \nu_{\pm} & = \frac{k}{N}  + \frac{1}{\pi} \arctan \biggl[  \frac{1}{\tan \frac{\pi}{N}} \biggl( 
   \frac{\eta +1}{\eta -1} \pm \\
   &\sqrt{ \left( \frac{\eta +1}{\eta -1}  \right)^2 + \tan ^2  \frac{\pi}{N} } \biggr) \biggr] ~,
    \end{split}
\end{equation}
where
\begin{equation}
   \eta = \frac{\chi_+ -1}{\chi_- -1} \qquad 
   \chi_\pm = \frac{|\phi(\nu_k)|^2}{|\phi(\nu_{k\pm1})|^2} \, .
\end{equation}
The choice between $\nu_-$ and $\nu_+$ depends on which of the coefficients $\vert \phi(\nu_{k-1}) \vert$ and $\vert \phi(\nu_{k+1}) \vert$ is the largest, i.e., $\nu_-$ is to be selected if 
\begin{equation}
    {\vert \phi(\nu_{k-1}) \vert > \vert \phi(\nu_{k+1}) \vert \, .}
\end{equation}

For the determination of the value of $\lambda$ one finds
\begin{equation}
\footnotesize
   \label{LambdaReconstruction_DampedExp}
   \lambda_{\pm} =
   2 \arcsinh \sqrt{ \frac{|\phi(\nu_{k\pm1})|^2
       \sin^2 \left( \pi \Delta \nu_k \mp \frac{\pi}{N} \right) -
       |\phi(\nu_{k})|^2 \sin^2 \pi \Delta \nu_k }
     {|\phi(\nu_{k})|^2 - |\phi(\nu_{k\pm1})|^2}      } \, ,
\end{equation}
where 
\begin{equation}
    {\Delta\nu_k = \nu - \nu _k \, ,}
\end{equation}
and the choice between $\lambda_-$ and $\lambda_+$ is made similarly to the case of the tune. 

Note that this approach provides an estimate of 
$\nu$ and $\lambda$ that are affected by an error that 
scales as $1/N^2$ 
(see Appendix~\ref{AppendixInterpolated_FFT_DampedExponential} 
for the details). 


In case the Hanning filter 
\begin{equation}
    {w (n) = 2 \sin^2 \left( \frac{\pi n}{N} \right)}
\end{equation}
is applied to the signal~\eqref{Definition_DampExp}, the {DFT} coefficients read 
\begin{equation}
   \label{phiNC_Hann}
\begin{split}
   \phi (\nu_j) & =
   \frac{1}{N} \sum_{n=1}^N e^{2 \pi i (\nu - \nu_j) n - \lambda n} w (n) \\
   & = \frac{2}{N} \sum_{n=1}^N e^{2 \pi i (\nu - \nu_j) n - \lambda n}
   \sin^2\frac{\pi n}{N} \, .
\end{split}
\end{equation}

By introducing the complex variable 
{
$$
\theta_j(\lambda) = \pi (\nu - \nu_j) + i \frac{\lambda}{2 },
$$
the Eq.}~\eqref{phiNC_Hann} can be simplified and analytically solved, providing a closed-form expression for $\nu$ and $\lambda$. A complex-valued quadratic equation is found in the unknown $\sin 2\theta_k$, where $k$ is again the index corresponding to the maximum of the amplitude of the Fourier spectrum, and its solutions are given by
\begin{equation}
   \label{LambdaEquation_Hanning_DampedExp}
   \lambda = \ln \left( \alpha + \sqrt{ 1+\alpha^2}\right) \, ,
\end{equation}
and
\begin{equation}
   \label{TuneEquation_Hanning_DampedExp}
   \nu = \frac{k}{N} + \frac{1}{2\pi} \arcsin \beta \, ,
\end{equation}
where the parameter $\alpha$ and $\beta$ are defined according to Eq.~(4.4.37) in Ref.~\cite{abramowitz}, and additional details on the mathematical derivation can be found in Appendix~\ref{Appendix_HanningDampedExponential}.

 Note that this approach provides an estimate of $\nu$ and $\lambda$ that are affected by an error that scales as $1/N^3$ (see Appendix~\ref{Appendix_HanningDampedExponential} for the details). 

\subsection{Hilbert Transform}
\label{Section_HilbertTransform}
\begin{figure}
    \centering
    \includegraphics[trim=40mm 31mm 52mm 43mm,width=\columnwidth,clip=]{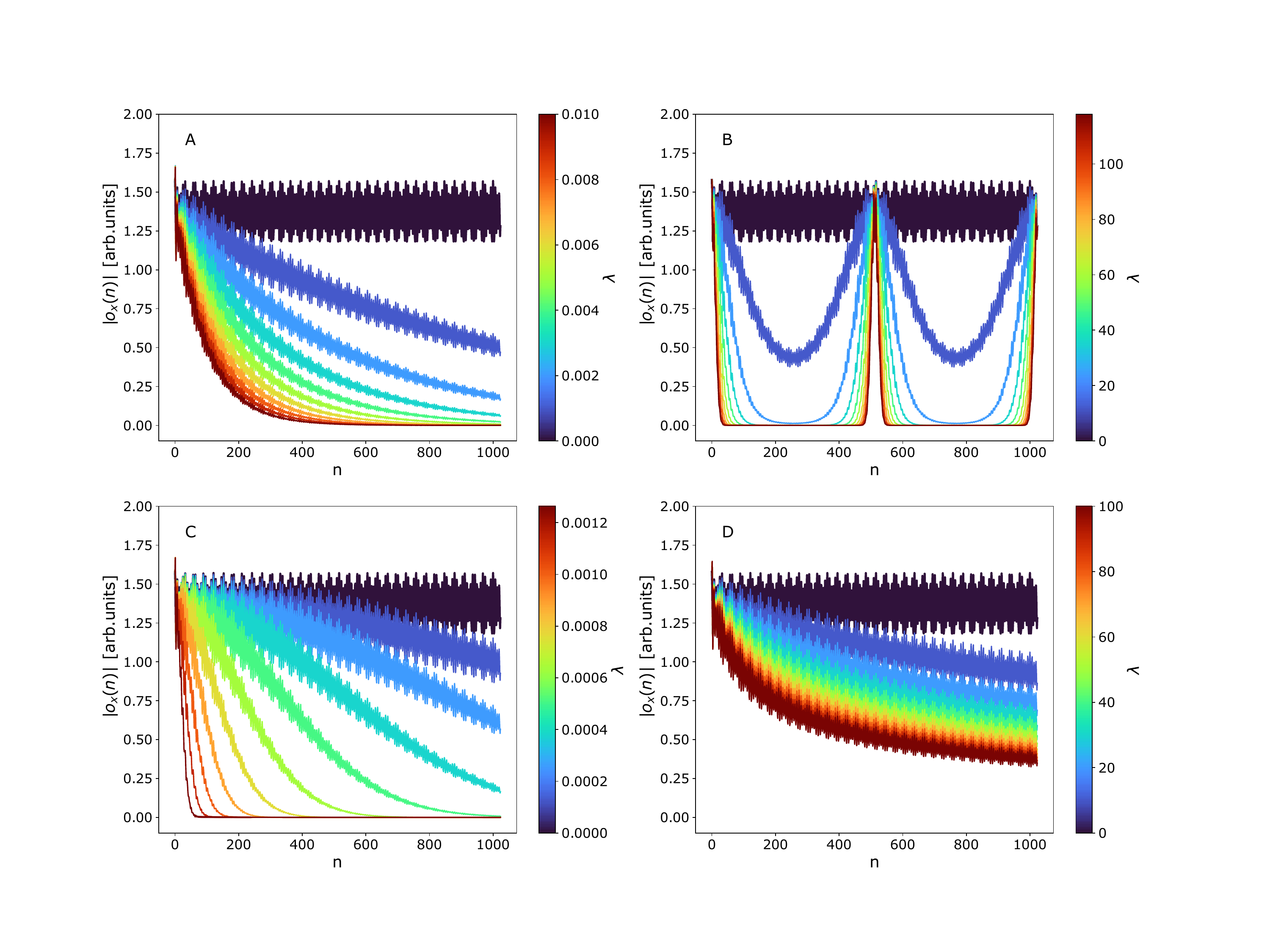}
    \caption{Reconstruction of the beam envelopes by means of the Hilbert transform for the signals with amplitude variation from energy damping (A), chromatic decoherence (B), nonlinearities (C), and acceleration (D) are shown as a function of the number of turns $n$ (up to the maximum of $1024$ turns). For each signal, the characteristic factor $\lambda$ has been varied in the interval shown on the color scale.} 
    \label{fig:ReconstructionEnvelopes}
\end{figure}

The Hilbert transform (HT) is one of the most powerful methods for determining the envelope of a signal. It is widely used in the processing of signals whose properties, such as amplitude, frequency, and statistical properties like mean, vary over time (see, e.g., \cite{Book_Hilbert, Paper_Hilbert}). HT has been used mainly in the analysis of real-valued signals, for this reason, we focus on real signals, and in this case the HT is defined as
\begin{equation}
   \label{HT_Definition}
   \mathcal{H}(x)(t) =
   - \frac{1}{\pi} p.v. \int_{-\infty}^{+\infty} \frac{x(\tau)}{t-\tau}~d\tau \, ,
\end{equation}
where $p.v.$ is the Cauchy principal value integral. In the case of a discrete-time signal, it is convenient to use the relationship between {DFT} and HT, which is given by
\begin{equation}
   \label{HilbertDefinition_FFT}
   \mathcal{H}(x)(n) = \mathcal{F}^{-1}( -i ~ sgn(\nu_\mathrm{{DFT}}) ~ \mathcal{F}(x(n)) ) \, ,
\end{equation}
with $\nu_\mathrm{{DFT}}$ the signal frequency vector, as determined with the {DFT}, $\mathcal{F}$ the {DFT} operator, and $\mathcal{F}^{-1}$ its inverse. 

It can be shown that the following complex-valued signal:
\begin{equation}
   \label{DefinitionAnalyticSignal}
   o_{x}(n) = x(n) + i \mathcal{H}( x(n) ) \, ,
\end{equation}
where the real part is the original time series and the imaginary part is the HT of the original time series, is an analytic signal. 

Consider a modulated signal of the form 
\begin{equation}
    {x(n) = u(n) x_0(n)} \, ,
\end{equation}
where $u(n)$ is a modulating function, and $x_0(n)$ a constant-amplitude signal with a single harmonic component of frequency $\nu$. Provided the frequency spectra of $u(n)$ and $x_0(n)$ do not overlap, and assuming that $u(n)$ contains the low-frequency and $x_0(n)$ the high-frequency component of the spectrum, then the theorem of Bedrosian~\cite{bedrosian:1963} can be applied and the HT of $x(n)$ is given by
\begin{equation}
    \mathcal{H}(x)(n)=u(n) \mathcal{H}(x_0)(n) \, , 
\end{equation}
which guarantees that $\vert o_x(n) \vert $ is the envelope of the analyzed signal, namely 
\begin{equation}\label{EnvelopeDefinitionHT}
    \begin{split}
  | o_x(n) | & = \vert u(n) \vert \, \vert  x_0(n) + i \mathcal{H}( x_0)(n) )  \vert \\
  & = \vert u(n) \vert \, \vert x_0(n) \vert \\
  & = \vert u(n) \vert \, .
    \end{split}
\end{equation}

By using the envelope, the original signal $x(n)$ can be normalized obtaining a constant-amplitude signal 
\begin{equation}
    \Tilde{x} (n) = \frac{x(n)}{\vert o_x(n) \vert}
\end{equation}
to which the methods to determine the tune developed in the past can be applied.

We note that the form of the prototype signal in Eq.~\eqref{SignalTune_OnlyRealPart} is very close to the factorization form needed by the Bedrosian theorem. The main difference is that the function $x(n)$ does not contain a single harmonic component, but rather a set of components. In this case, assuming that the conditions on the spectra of the two factors of the signal are satisfied, we obtain:
\begin{equation}
\begin{split}
      o_x(n) & = a e^{f(\lambda, n)} \left [ \left( \cos 2\pi n \nu + \sum_{k=1}^4 a_k \cos 2\pi n k \nu \right ) + \right . \\
      & \left . + i \left( \sin 2\pi n \nu + \sum_{k=1}^4 a_k \sin 2\pi n k \nu \right ) \right ] \, , 
\end{split}
\end{equation}
by applying the properties of the HT to the harmonic components of the function representing $x(n)$. In this case, $\vert o_x(n)\vert $ is not exactly the envelope of the function $x(n)$ but reads
\begin{equation}
    \begin{split}
    \vert o_x (n) \vert & = a e^{f(\lambda, n)} \left ( 1+2 \sum_{k=1}^4 a_k \cos 2\pi n (k-1) \nu + \sum_{k=1}^4 a_k^2 + \right . \\
    & \left . + 2 \sum_{k=1, j> k}^4 a_k a_j \cos 2\pi n (k-j) \nu \right )^{1/2} \\        
    & \approx a e^{f(\lambda, n)} \left ( 1 + \sum_{k=1}^4 a_k \cos 2\pi n (k-1) \nu \right )
    \end{split}
    \label{eq:normalization_develpoment}
\end{equation}
where the last step assumes that 
\begin{equation}
    {a_k \ll 1} \, ,
\end{equation}
i.e., the harmonic components are indeed perturbations of the main component. The normalized signal then reads
\begin{equation}
    \begin{split}
    \Tilde{x} (n) & \approx \left ( \cos 2\pi n \nu + \sum_{k=1}^4 a_k \cos 2\pi n k \nu \right ) \times \\
    & \times \left ( 1 - \sum_{k=1}^4 a_k \cos 2\pi n (k-1) \nu \right ) \\
    & \approx \cos 2\pi n \nu + \frac{1}{2} \sum_{k=1}^4 a_k \cos 2\pi n k \nu + \\
    & -\frac{1}{2} \sum_{k=1}^4 a_k \cos 2\pi n (k-2) \nu \, .
    \end{split}
    \label{eq:normalized_development}
\end{equation}

The signal $\Tilde{x}(n)$ contains the same frequencies as the original signal, but the amplitudes are affected by the application of HT. The situation would be even more involved if the original signal were made by a spectrum consisting of several different frequencies, not being harmonics of a single frequency, as in this case the equations above predict the generation of new spurious frequencies in the normalized signal. A similar situation would occur in the case where the perturbative approach cannot be used and higher-order terms should be added to Eqs.~\eqref{eq:normalization_develpoment} and~\eqref{eq:normalized_development}.

Note also that the feasibility of reconstructing the function $u(n)$ can be seen as an a posteriori validation of the applicability conditions of the Bedrosian theorem. 

All these considerations confirm that the approach based on the HT is a very powerful tool for performing harmonic analysis of signals with varying amplitude. In fact, it allows methods such as interpolated {DFT} and interpolated {DFT} with the Hanning filter to be applied in this case as well. Furthermore, it provides a means to reconstruct the envelope of the signal, which contains essential information to describe the physical phenomenon that generates the variation of the amplitude. This aspect will be used in the following to propose an alternative method for measuring the amplitude detuning that will be applied to the analysis of the LHC data in Section~\ref{DecoherenceProtonBeams_Section}. 

\section{Accuracy studies}\label{NumericalSimulations}

In this section, we present the numerical studies performed to assess the accuracy of the proposed methods for reconstructing the tune of amplitude-varying signals. For every amplitude modulation model discussed in Section~\ref{sec:models}, the tune is computed by applying the Hilbert transform to normalize the original signal and then using the two standard methods, the interpolated {DFT} and the interpolated {DFT} with Hanning filter, applied to the normalized signal. 

A special analysis will be performed on the signal describing the energy-damping case, as the closed-form solution can be applied here to determine the tune $\nu$ and the damping factor $\lambda$. 

\begin{figure}
    \centering
    \includegraphics[trim=1mm 4mm 16mm 16mm,width=0.6\columnwidth,clip=]{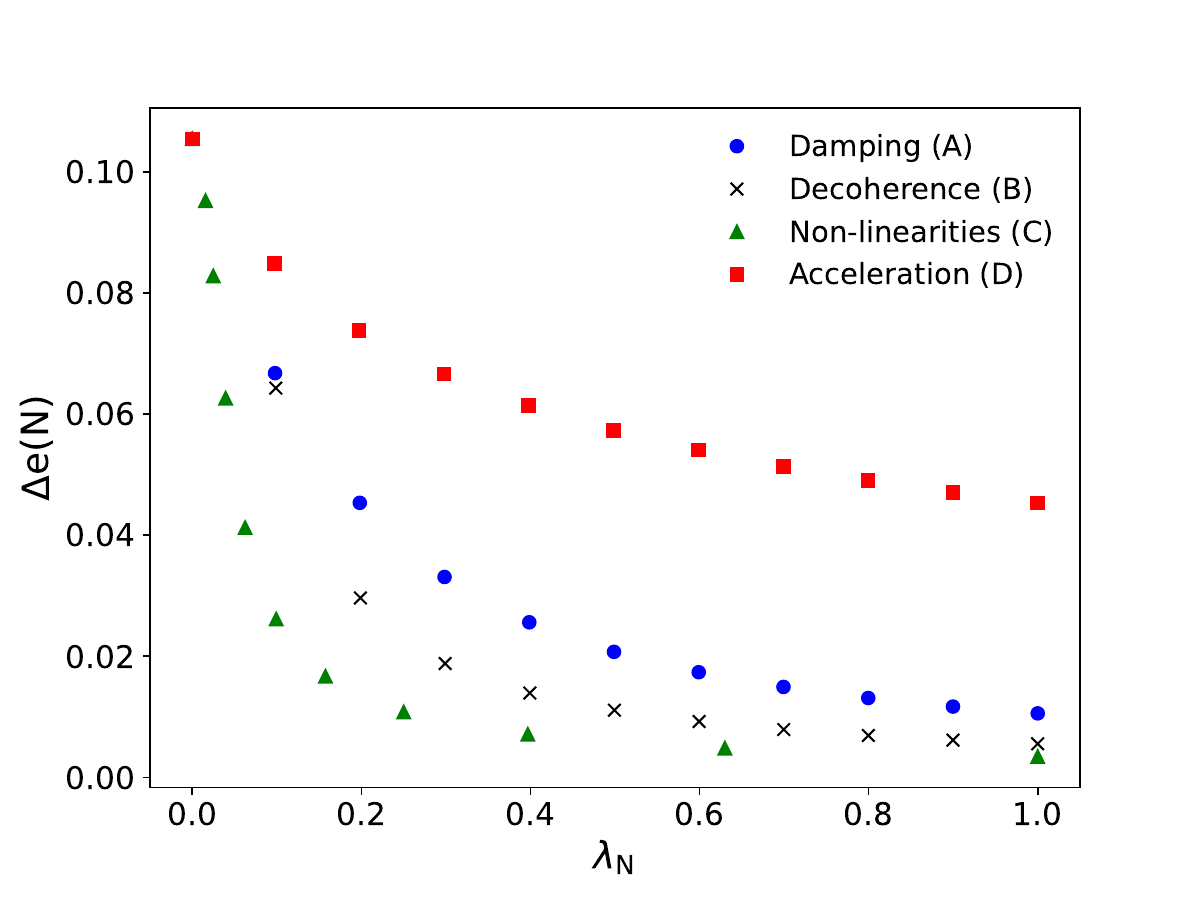}
    \caption{{Mean Square Error of} the reconstruction of the signal envelopes reported in Fig.~\ref{fig:ReconstructionEnvelopes} as a function of the parameter $\lambda_\mathrm{N}$, normalized in the interval $[0,1]$.} 
    \label{fig:AccuracyReconstructionEnvelopes}
\end{figure}

\begin{figure}
    \centering
    \includegraphics[trim=65mm 2mm 68mm 15mm,width=\columnwidth,clip=]{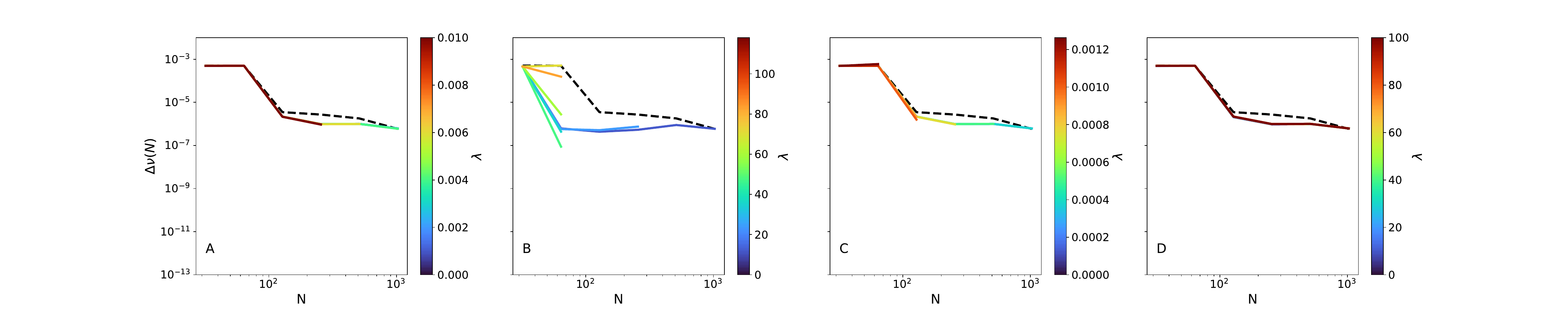}
    \includegraphics[trim=65mm 2mm 68mm 15mm,width=\columnwidth,clip=]{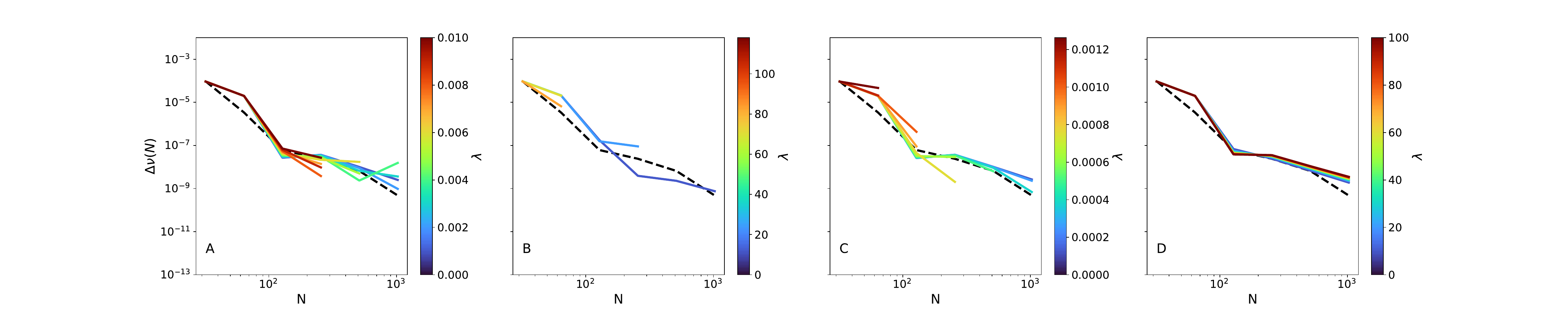}
    \caption{The tune error $\Delta \nu(N)$ is computed for the signals with amplitude variation from energy damping (A), chromatic decoherence (B), nonlinearities (C), and acceleration (D) (from left to right). Two methods have been used, the interpolated {DFT} (top) and the interpolated {DFT} with Hanning filter (bottom), and the tune is computed after having normalized the signal using the Hilbert transform. {The scaling of the tune error corresponding to the case $\lambda=0$ is represented by the back dashed lines with a dependence as $1/N^2$ (top) and $1/N^4$ (bottom).} In general, one observes that when the HT is applied, the original performance of the tune determination using the {DFT}-based methods is restored, i.e., it is independent on the value of $\lambda$.} 
    \label{fig:TuneErrorAllCaseStudies}
\end{figure}

First, it is necessary to verify whether the use of the HT to normalize the signal is justified according to the conditions stated in the Bedrossian theorem. Figure~\ref{fig:ReconstructionEnvelopes} shows the envelopes of the four modulated amplitude signals that are used as prototypes in our study. Key information is shown in Fig.~\ref{fig:AccuracyReconstructionEnvelopes}, where the error in the reconstruction of the signal envelope as a function of $\lambda$ is visible for the four signals. As an estimate of the error, we used the quantity
\begin{equation}
    \Delta e(N) = \frac{1}{N} \sum_{n=1}^N \left ( \frac{\vert o_x(n)\vert -\vert o^\mathrm{model}_x(n) \vert}{\vert o_x(n) \vert} \right )^2 \, , 
\end{equation}
{which corresponds to the Mean Square Error,} where $\vert o_x(n) \vert$ and $\vert o^\mathrm{model}_x(n)$ are the signal envelope reconstructed using the HT and that computed using the physical model, respectively. 
Globally, the quality of the envelope reconstruction is very good, which confirms the applicability of the Bedrossian theorem and the fact that it is possible to determine the tune of the normalized signal with HT by using the interpolated {DFT} with or without the Hanning filter. For each amplitude-modulated signal, the tune error is displayed as a function of $N$, the number of samples in the signal. Note that for a given value of $\lambda$, $N$ is limited to values for which the signal amplitude is approximately not less than $10\%$ of its initial value. This choice agrees with the common practice employed when analyzing real turn-by-turn beam data. Therefore, signals with a strong modulation of amplitude, corresponding to high values of $\lambda$, are analyzed using a shorter value of $N$. The results of the numerical simulations for the interpolated {DFT} (top) and the interpolated {DFT} with Hanning filter (bottom) are shown in Fig.~\ref{fig:TuneErrorAllCaseStudies}. {For comparative purposes, the dashed lines depict the curves $1/N^2$ (top) and $1/N^4$ (bottom), which illustrate the scaling of the tune error for the benchmark methods, specifically the Interpolated DFT and the Interpolated DFT with Hanning filter, respectively.}
%
\subsection{Amplitude variation from energy damping}\label{NumericalResultsExponentialDamping}

The results of the application of the HT method are reported in Fig.~\ref{fig:TuneErrorAllCaseStudies}~(A), where the tune error is shown when using the interpolated {DFT} (top) and the interpolated {DFT} with Hanning filter (bottom). Faster error reduction is observed, as expected, for the interpolated {DFT} with the Hanning filter. Moreover, the tune error is almost independent of $\lambda$ for the case of interpolated {DFT}, whereas a variation is visible for the use of interpolated {DFT} with Hanning filter. 

\begin{figure}
    \centering
    \includegraphics[trim=37mm 2mm 49mm 15mm,width=0.6\columnwidth,clip=]{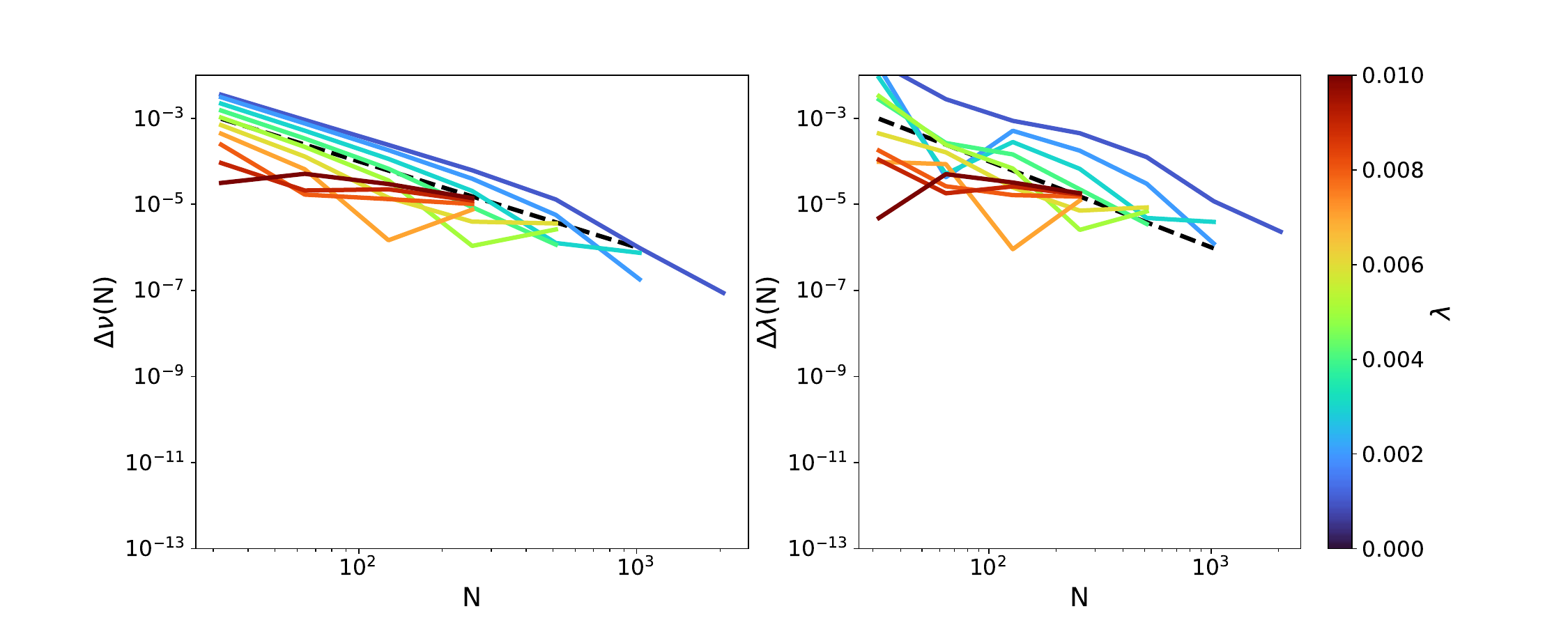}
    \includegraphics[trim=37mm 2mm 49mm 15mm,width=0.6\columnwidth,clip=]{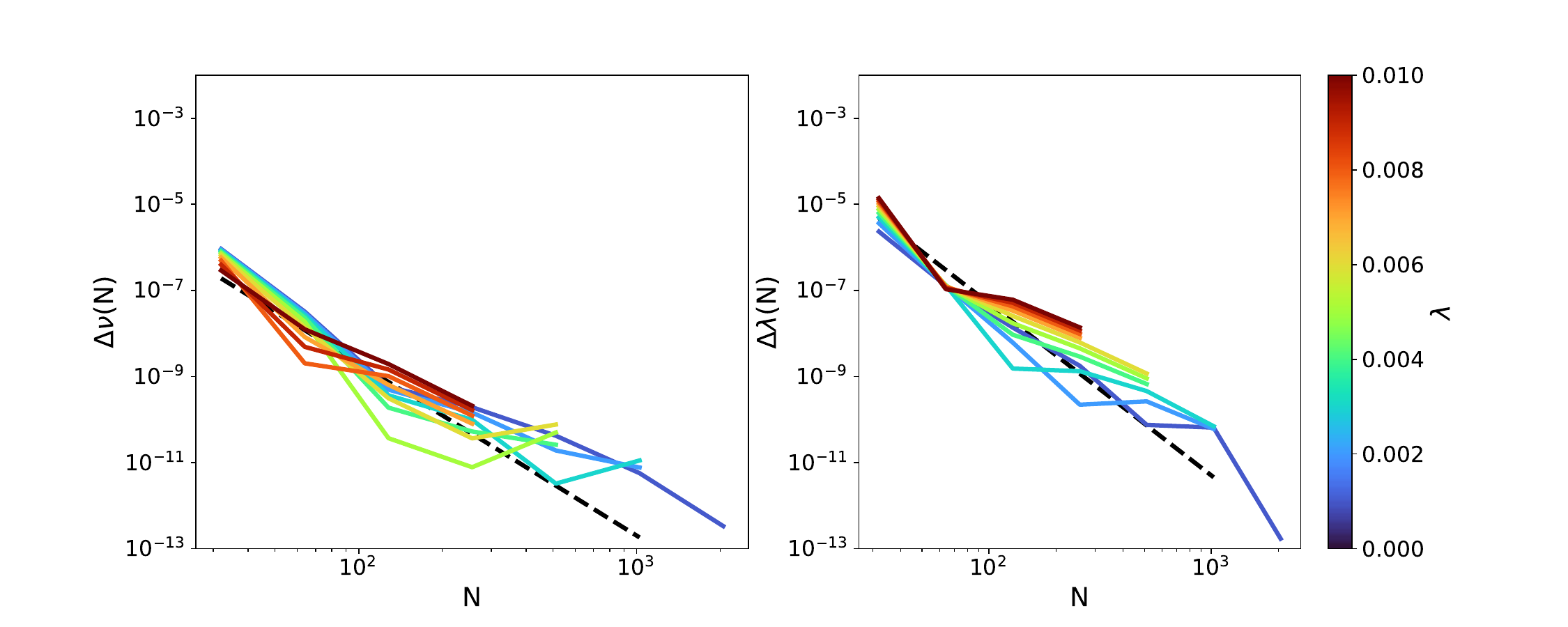}
    \caption{Dependence on $N$ of the error on $\nu$ (left) and $\lambda$ (right) for the damped exponential signal using the interpolated {DFT} ( Eqs.~\eqref{Tune_DampedExponential} and~\eqref{LambdaReconstruction_DampedExp}, top) and interpolated {DFT} with Hanning filter (Eqs.~\eqref{TuneEquation_Hanning_DampedExp} and~\eqref{LambdaEquation_Hanning_DampedExp},
 bottom) for several values of $\lambda$. {The expected scaling of the tune error is represented by the back dashed lines with a dependence as $1/N^2$ (top) and $1/N^4$ (bottom). The colored curves are all featuring a slope in agreement with the expectations}.} 
    \label{fig:DampingFactorExponential}
\end{figure}

The damped exponential is the only type of signal for which there is an exact equation to determine the tune and damping factor, and the results of the application of the equations derived in Section~\ref{sec:closed-form-damped} are shown in Fig.~\ref{fig:DampingFactorExponential}. {For comparative purposes, the dashed lines depict the curves $1/N^2$ (top) and $1/N^4$ (bottom), which illustrate the scaling of the tune error for the benchmark methods, specifically the Interpolated DFT and the Interpolated DFT with Hanning filter, respectively.} The two methods provide a tune error that scales as $1/N^2$ for the interpolated {DFT} and $1/N^4$ for the interpolated {DFT} with the Hanning filter. 

The interesting feature of the exact equations is the possibility of also finding a solution for the damping parameter $\lambda$. The error of the reconstructed values of $\lambda$ scales are $1/N^{2}$ and $1/N^{4}$ for the interpolated {DFT} and the interpolated {DFT} with Hanning filter, respectively.

In the reconstruction of $\lambda$, {it is observed that the error on $\lambda$ saturates beyond a certain value $N$. This is due to the exponential damping of the signal amplitude: beyond a certain value of $N$ the signal is compatible with zero with the double precision accuracy of the numerical computations.} 
%
\subsection{Amplitude variation from chromatic decoherence}\label{NumericalSimulation_ChromDecoherence}

The chromatic signal described in Eq.~\eqref{ChromaticBeamCentre} has an amplitude modulation that depends on the number of turns according to a sinusoidal function. This feature makes it impossible to find a closed-form solution of the equation of the {DFT} coefficient for the tune. The tune of the chromatic signal can be determined by exploiting Hilbert normalization and the methods described above. The results are shown in Fig.~\ref{fig:TuneErrorAllCaseStudies}~(B). In general, the results confirm that normalization by means of HT restores the accuracy of the methods based on {DFT} for signals of constant amplitude.

This type of signal has been used to assess the accuracy in determining the physical parameters from the knowledge of the signal envelope reconstructed by means of the HT technique. 

\begin{figure}
    \centering
    \includegraphics[trim=0mm 2mm 10mm 15mm,width=0.6\columnwidth,clip=]{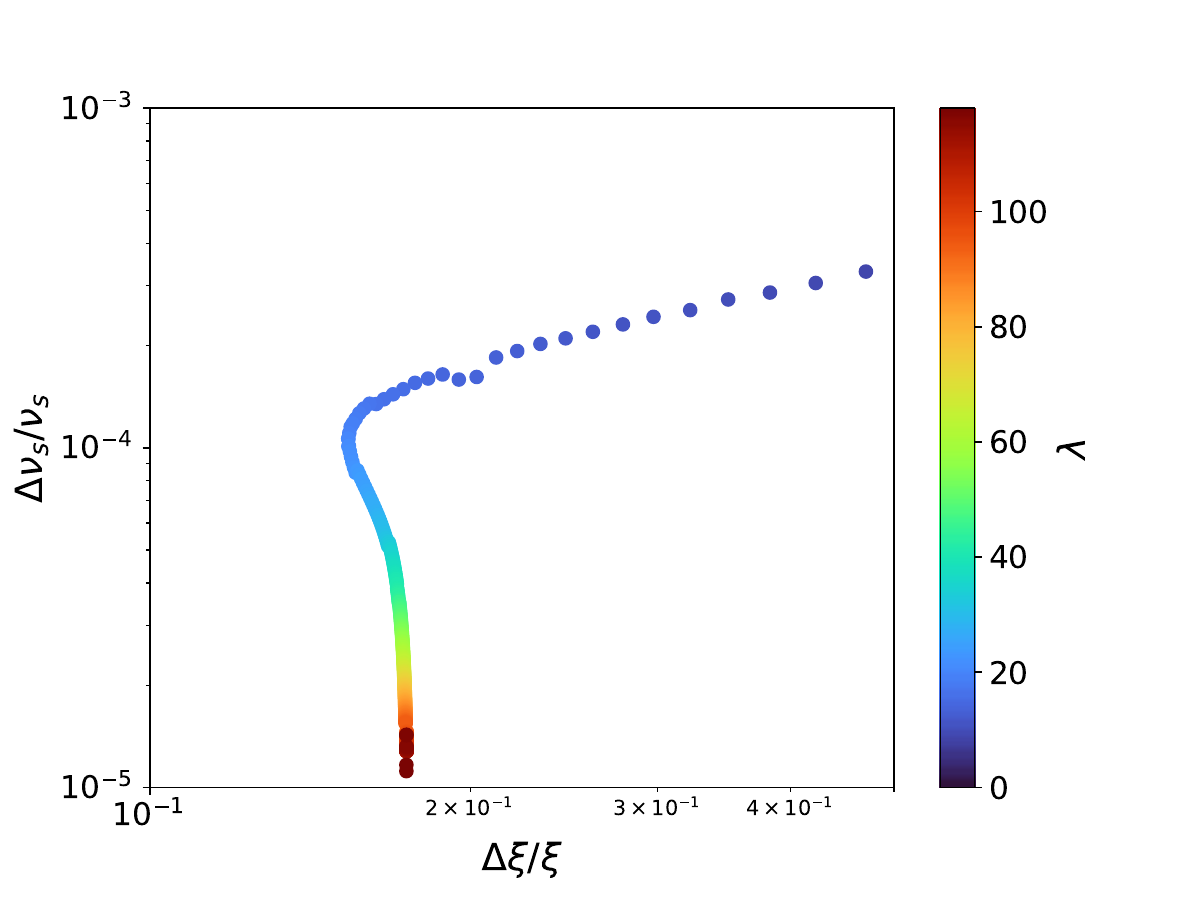}
    \caption{Relative error of the reconstructed values of $\nu_\mathrm{s}$ and $\xi$ from the fit of the signal envelope obtained using the HT approach. The color code represents the value of the damping factor $\lambda$. The large relative error associated with small chromaticity values highlights the difficulty in reconstructing the parameter when the modulation of the amplitude due to chromatic decoherence is relatively small, namely below $10\%$.} 
    \label{fig:ErrorReconstructionChromaticSignal}
\end{figure}

The modulation of the amplitude described by the exponential term of Eq.~\eqref{ChromaticBeamCentre} has been fitted to the envelope. The three physical parameters, namely synchrotron tune $\nu_{\mathrm{s}}$, chromaticity $\xi$, and rms {relative momentum spread} $\sigma_\delta$, can be used as model fit parameters in various combinations. The experimental determination of $\sigma_\delta$ is not difficult, as it can be measured quite easily. In principle, the synchrotron tune can be derived from knowledge of the fundamental machine parameters. However, it is more critical to determine the chromaticity, as it requires measuring the tune for various values of the momentum offset and the synchronous tune. The approach consisted of studying the best pair of parameters to fit the envelope. Calculating $\xi$ and $\sigma_\delta$ from the fit proved to be quite challenging, and the associated error was large. In fact, in Eq.~\eqref{LambdaChromaticDecoherence}, $\xi$ and $\sigma_\delta$ are multiplied by each other, so the fit tends to optimize their product rather than the single variable. If $\nu_{\mathrm{s}}$ and $\sigma_\delta$ are used as fit parameters, large errors are observed in the synchrotron tune (approximately $\approx 40\%$ for the configuration where no recoherence occurs within the first $1024$ turns). Therefore, the best set of parameters for the fit procedure was that of $\xi$ and $\nu_{\mathrm{s}}$ and the results are reported in Fig.~\ref{fig:ErrorReconstructionChromaticSignal}. 

The error associated with the reconstruction of the synchrotron tune is reduced as the value of chromaticity increases. In fact, the longer the time interval of the re-coherence phenomenon, for a fixed number of turns, the more precise the reconstruction of $\nu_{\mathrm{s}}$, as it is closely related to the phenomena of chromatic decoherence and re-coherence. As $\xi$ increases, a larger error is observed, but it is still always below $10\%$. {To summarize, this method appears to be an effective way to ascertain fundamental machine parameters like chromaticity and synchrotron tune.}

\subsection{Amplitude variation from nonlinearities and acceleration}\label{TuneErrorAMplitudeDetuningNumericalSimulation}

The damping of coherent oscillations caused by nonlinearities and acceleration is characterized by amplitude modulations that cannot be treated using a direct {DFT}-based method because no exact closed-form solution can be found. Therefore, the HT-based approach is the only remaining option. 

The results are shown in Figs.~\ref{fig:TuneErrorAllCaseStudies}~(C) and (D) for the case of signals with nonlinearity and acceleration, respectively. The efficient reconstruction of the tune value is clearly visible, with an accuracy that is, to a large extent, independent of the value of the damping parameter $\lambda$. {Thus, employing the HT-based method for tune determination, which is the sole suitable technique in these scenarios, has proven to be highly precise, reinstating the efficiency of the high-accuracy tune determination methods developed for stationary signals.}

\section{Application to LHC data}\label{DecoherenceProtonBeams_Section}

The promising results discussed in the previous sections suggested the application of the HT-based approach to beam data collected at the LHC. 

During the 2012 experimental campaign, an attempt was made to measure the detuning with amplitude of the LHC lattice using kicked beams~\cite{EwenPaper}. The data had already been analyzed using the standard approach, relying on harmonic analysis of the raw data, which includes a non-negligible variation of the signal amplitude (see Fig.~\ref{fig:Example_LHC_signal} for an example). A new analysis could be based on the use of the HT to normalize the original data and then perform a harmonic analysis of the normalized data. 

We note here that the signal decoherence model that corresponds to the LHC measurements is the one discussed in Section~\ref{NonLinearDecoherence} and presented in Ref.~\cite{Meller:1987ug}, which relies on two main assumptions, namely that the transverse beam distribution is Gaussian and the amplitude detuning is a quadratic function of the betatron amplitude (hence, linear in the action variable). Therefore, the envelope information of the signal could be used to fit the model parameters and hence determine the amplitude detuning from each single turn-by-turn measurement. In fact, we stress that, unlike the standard approach to determine the amplitude detuning, which consists of displacing the beam at various amplitudes and determining the tune at each amplitude to reconstruct the dependence of the tune from the amplitude, the proposed approach is capable of reconstructing the amplitude detuning with a single measurement. The potential gain of the new approach is clearly apparent.

The amplitude detuning is defined as the Taylor expansion of the tune according to:
\begin{equation}
    \begin{split}
 Q_u (J_x, J_y) & = Q_{u, 0} + \frac{\partial Q_u}{\partial J_x} J_x + \frac{\partial Q_u}{\partial J_y} J_y + \\
& + \frac{1}{2} \left (   \frac{\partial^2 Q_u}{\partial J_x ^2} J_x^2 +
    \frac{\partial^2 Q_u}    {\partial J_x \partial_y J_y} J_x J_y +
    \frac{\partial^2 Q_u}{\partial J_y^2} J_y^2 \right ) \, , 
    \end{split}
    \label{Tune_TaylorExpansion}
\end{equation}
where $u=x, y$ and $J_u$ stands for the single-particle linear action. Note that {the action is not normalized to beam size, and the amplitude detuning coefficients $\mu$ and $\mu_2$ introduced later are not dimensionless as for the case of model~C in Section~\ref{NonLinearDecoherence}. Furthermore, it should be noted that} the coefficients in Eq.~\eqref{Tune_TaylorExpansion} are not all independent, but satisfy the following relationships
\begin{align}
\frac{\partial Q_x}{\partial J_y} & = \frac{\partial Q_y}{\partial J_x} \\
\frac{\partial^2 Q_x}{\partial J_y ^2} & = 2     \frac{\partial^2 Q_y}{\partial J_x \partial J_y}   \qquad  \frac{\partial^2 Q_y}{\partial J_x ^2} = 2     \frac{\partial^2 Q_x}{\partial J_x \partial J_y} \, .
\end{align}

Amplitude detuning is measured by kicking the beam and recording the position of the beam centroid on a turn-by-turn basis. At the end of the measurement, a signal sampled over $N$ turns is available for analysis. {DFT}-based techniques, such as NAFF~\cite{Laskar:2003a} or SUSSIX~\cite{Bartolini:1997np,Bartolini:702438}, are used to determine the tune for each of approximately 500 LHC BPMs, and the average, over the BPMs, of the tune values provides an estimate of the tune, while the standard deviation provides an estimate of the associated tune error due to the reproducibility of the beam measurement. When the measurement and frequency analysis are repeated for different kick amplitudes, it is possible to obtain the tune as a function of the single-particle action. A fit is then performed on the experimental tune data to determine the form of the tune variation, and thus the partial derivative terms in Eq.~\eqref{Tune_TaylorExpansion} are determined. 

\subsection{Comparison with 2012 measurements results}\label{NonLinearSignal_subsection}

\begin{figure}[htb]
    \centering
    \includegraphics[trim=6mm 6mm 3mm 3mm,width=0.6\columnwidth,clip=]{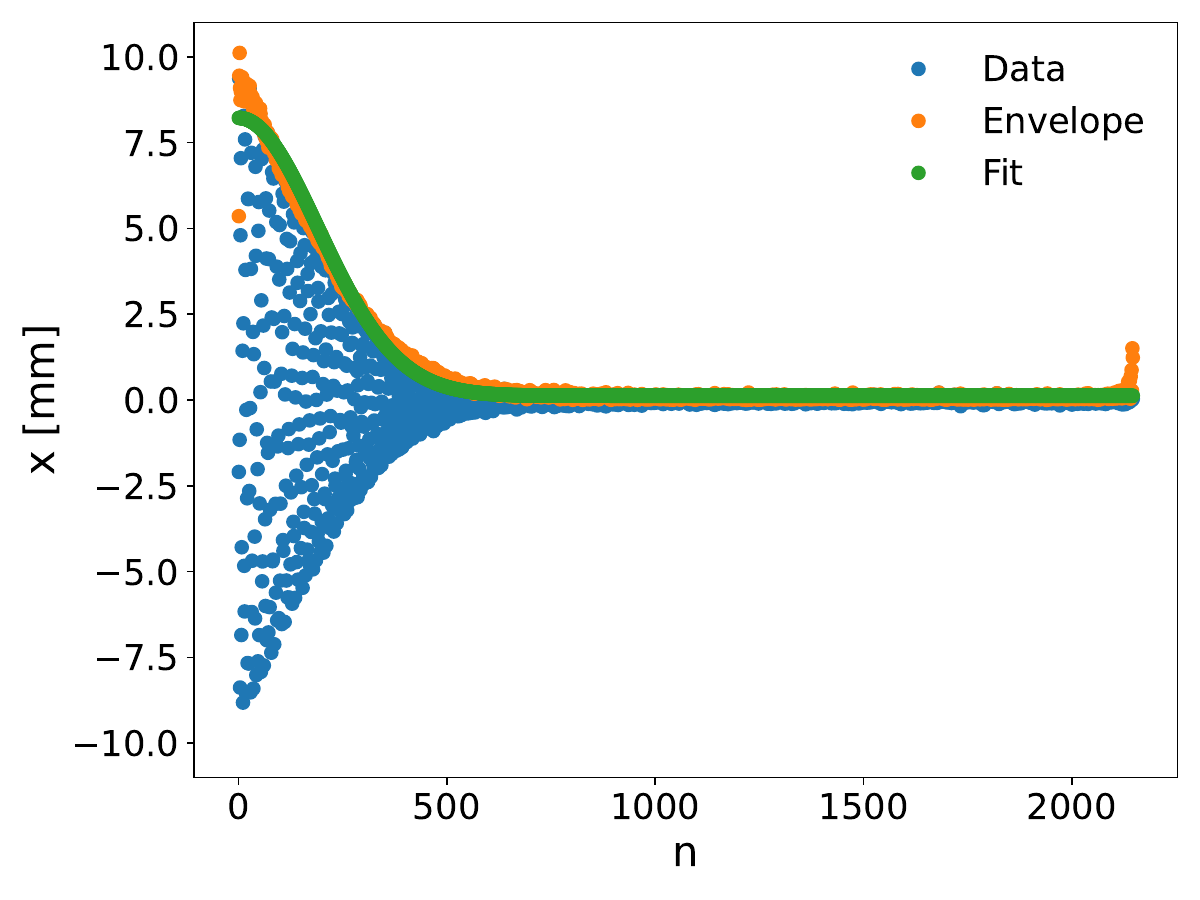}
\caption{Turn-by-turn BPM measurement of the beam centroid (black). The BPM is located in cell 5 to the left of the ATLAS interaction point and provides the horizontal position of the counterclockwise beam. The envelope computed with the HT-based approach is also shown (orange), together with the result of the fit procedure (green) based on the model describing the decoherence in the presence of nonlinearities.}
\label{fig:Example_LHC_signal}
\end{figure}

An example of a BPM signal acquired during 2012 LHC measurements is shown in Fig.~\ref{fig:Example_LHC_signal}, where a clear {reduction of the signal} amplitude is visible, which confirms that, based on previous considerations, {conventional Fourier transform analysis as executed by the NAFF software}~\cite{Laskar:2003a} is not applicable. 

The new analysis is based on the following steps: The envelope of every BPM signal is reconstructed by HT and used to normalize the signal. Then, the tune is calculated using the interpolated {DFT} on the constant-amplitude signal. Finally, the envelope is fitted with a Gaussian function, namely
    \begin{equation}\label{FitGaussianEnvelope}
        \vert o_\mathrm{Fit} (n) \vert = A e^{-\lambda n^2}\, ,
    \end{equation}
where the factor $\lambda$ and the amplitude $A$ are the fit parameters. In Fig.~\ref{fig:Example_LHC_signal}, the reconstructed envelope is shown in orange, while the result of the BPM fit procedure is shown in green. 

During the 2012 measurement campaign, the amplitude detuning was measured for two LHC configurations: An uncorrected configuration, where no adjustment of the lattice nonlinear corrector magnets was performed; a corrected configuration, where the strengths of specific lattice nonlinear corrector magnets were adjusted to minimize the effects of nonlinear field errors, which was used in operations (additional details can be found in Ref.~\cite{EwenPaper}).
\begin{figure}
    \centering
    \includegraphics[trim=28mm 30mm 40mm 42mm,width=\columnwidth,clip=]{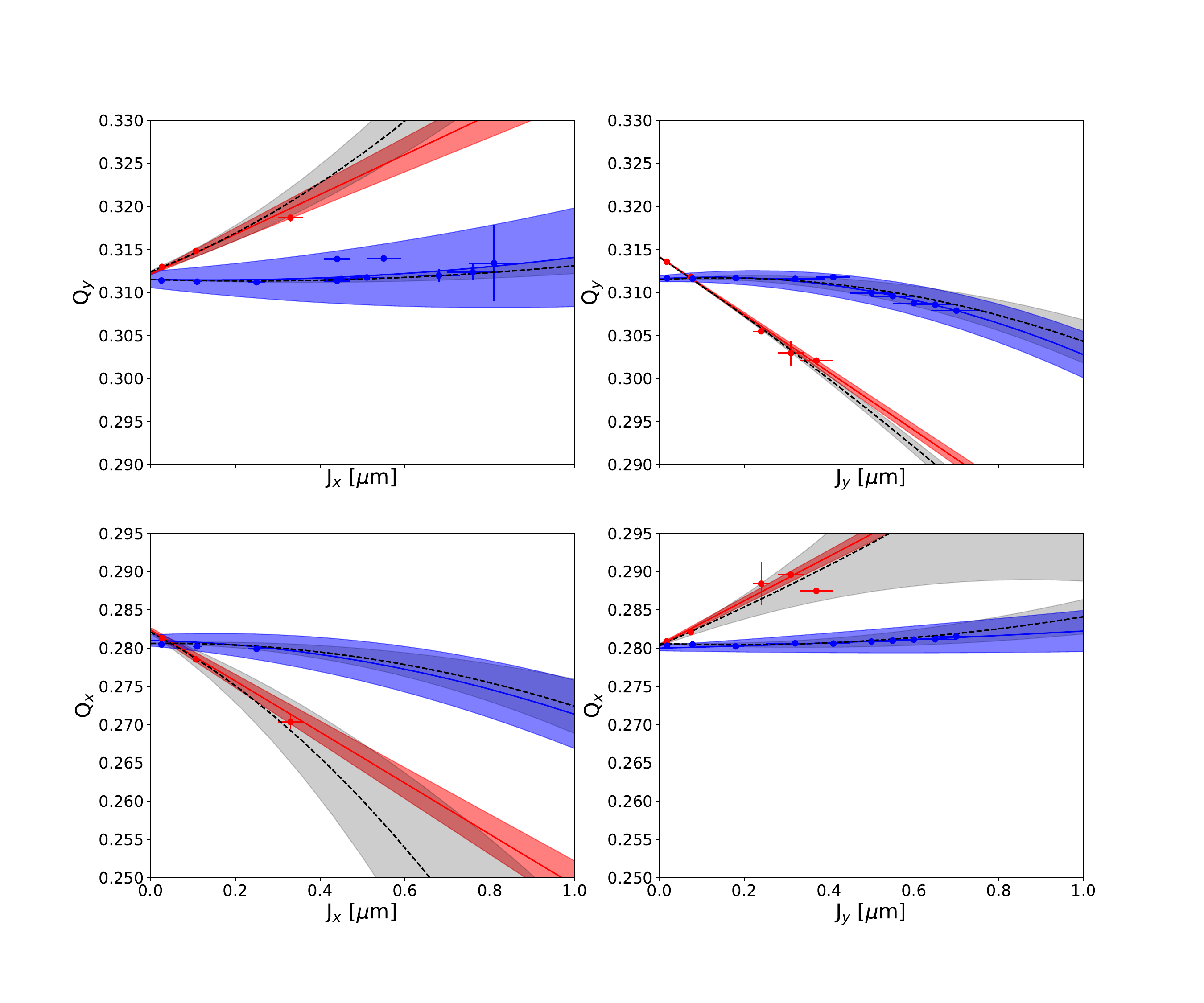}
\caption{The data collected during the 2012 experimental campaign at LHC have been analyzed using the HT-based approach. The red data and the corresponding fit refer to the uncorrected LHC configuration, whereas the black data refer to the corrected {LHC} configuration. The results obtained in 2012 with the standard analysis are shown in black and gray. The picture shows the change of the vertical (top row) and horizontal (bottom row) tunes as functions of the horizontal (left column) and vertical (right column) linear actions. The colored bands represent the rms error from the fit procedure.}
\label{fig:AmplitudeDetuningPlot_ComparisonWithEwen}
\end{figure}

The results of the new analysis performed on the two data sets are shown in Fig.~\ref{fig:AmplitudeDetuningPlot_ComparisonWithEwen}, where the uncorrected configuration is shown in red, and the corrected configuration is shown in black. The plots show the vertical (top) and horizontal (bottom) tune as a function of a horizontal (left) and vertical (right) single-particle linear action. Analysis of the experimental data includes error bars for the tune and action, respectively. 

The measured data are also fitted with
\begin{equation}\label{Fit_AmplitudeDetuning_Tune}
    Q_u (J_{u^\prime}) = Q_{u,0} + \mu J_{u^\prime} + \frac{1}{2} \mu_2 J_{u^\prime}^2\, .
\end{equation}
where the subscripts $u$ and $u^\prime$ stand for $x, y$ in all possible combinations and $\mu$ has the same meaning as the parameter of Eq.~\eqref{eq:amplitude-detuning}, while $\mu_2$ represents a second-order detuning term that is not present in the derivation of {the model describing the amplitude variation due to nonlinearities of Eq.}~\eqref{NonLinearDecoherenceBeamCentroid}. The result of the linear ($\mu_2 = 0$) or parabolic fit from Eq.~\eqref{Fit_AmplitudeDetuning_Tune} is shown in Fig.~\ref{fig:AmplitudeDetuningPlot_ComparisonWithEwen} with a continuous line, and the rms error of the fit procedure is shown with the colored area. The dashed black curve and the corresponding gray error area refer to the fit and its error obtained in 2012 with the previous approach (see Ref.~\cite{EwenPaper} for more details). 

When comparing the results obtained with the two approaches, i.e., the standard approach (black lines and gray areas) and the new approach (colored lines and colored areas), it is possible to observe general agreement. The uncorrected configuration (in red) shows a reduction in the fit uncertainty. A visible discrepancy is observed in the uncorrected case for the variation in the tune as a function of $J_x$ (left plots). {It is worth mentioning that} when the current analysis is compared with that presented in Ref.~\cite{EwenPaper}, a quadratic term is missing in the new data. This is the consequence of dropping a couple of experimental points a large value of $J_x$. This choice is based on the observation that at the largest amplitudes the beam is trapped in stable islands. Hence, the measured tune value is not representative of the natural amplitude detuning. measured tune value. As far as the variation of the tune as a function of $J_y$ is concerned (plots in the right column), the agreement between the fitted lines {(red and black for the new analyses and black for the 2012 analyses)} is much better, with a slope that agrees at the level of $\approx 7\%$, which is within the $10\%$ {level that is commonly assumed as the measurement accuracy of this type of beam dynamics measurements at the LHC}. 

As far as the corrected configuration (black curve) is concerned, the uncertainty associated with the fit is larger than that obtained from the previous method. However, the fit curve represents the experimental data well, which is also confirmed by the value of the reduced chi-square test. The coefficients in Eq.~\eqref{Fit_AmplitudeDetuning_Tune} and the corresponding reduced $\chi^2$ are reported in Table~\ref{Table_ResultsFit_AmplitudeDetuning}.
\begin{table*}[htb]
    \centering
    \caption{Results of the fit of the tune as a function of the single-particle action {using Eq.~\eqref{Fit_AmplitudeDetuning_Tune}}.}
    \begin{tabular}{c|c|c|c|c}
    \toprule
            & \text{$Q_y$ \text{vs} $J_x$} & \text{$Q_y$ \text{vs} $J_y$}& \text{$Q_x$ \text{vs} $J_x$} & \text{$Q_x$ \text{vs} $J_y$}  \\
    \cline{2-5} 
            & \multicolumn{4}{c}{Uncorrected {LHC} configuration}  \\
    \cline{1-5} 
  $Q_{u,0}$                          & $0.3121 \pm 0.0001$ & $0.31419 \pm 0.00004$ & $0.2824 \pm 0.0003$ & $0.2804 \pm 0.0002$ \\
  $\mu  $ [\SI{e4}{\micro m^{-1}}]  & $2.3 \pm 0.3$        & $-3.4 \pm 0.1$        & $-3.3 \pm 0.3$        & $2.9 \pm 0.2$            \\ 
  $\mu_2$ [\SI{e10}{\micro m^{-2}}] & $-$        & $-$                   & $-$        & $-$  
                      \\ 
  $\chi^2 [\SI{e-2}{}]$             &   $2$                & $3$                   & $1$                   & $0.02$  
                      \\ 
    \midrule
    \text{} & \multicolumn{4}{c}{Corrected {LHC} configuration}  \\
    \cline{1-5} 
  $Q_{u,0}$                          &  $0.3115 \pm 0.0009$& $0.3116 \pm 0.0003$   & $0.2810 \pm 0.0007$   &   $0.2800 \pm 0.0003$ \\
  $\mu$ [\SI{e4}{\micro m^{-1}}]     & $-0.9 \pm 0.5$           & $0.3 \pm 0.2$        & $-0.1 \pm 0.4$        & $0.1 \pm 0.2$            \\ 
  $\mu_2$ [\SI{e10}{\micro m^{-2}}]  & $0.70771 \pm 0.00001$     & $-2.35561 \pm 0.00007$       & $-1.69195 \pm 0.00001$        & $0.17039 \pm 0.00001$           \\ 
  $\chi^2 [\SI{e-2}{}]$              & $0.02$              & $0.05$                & $0.002$               & $0.002$                   \\ 
    \bottomrule
    \end{tabular}\\
    \label{Table_ResultsFit_AmplitudeDetuning}
\end{table*}

From the results reported in Table~\ref{Table_ResultsFit_AmplitudeDetuning}, the largest uncertainty is related to the determination of the first-order coefficient $\mu$ for the case of the corrected configuration, but this is a consequence of the reduction in the first-order detuning due to the correction procedure. As a consequence, the tune dependence is strongly dominated by the second-order term, and this makes the determination of the first-order term difficult and, in all the cases, the values are compatible with zero. However, the low value of the reduced $\chi^2$ confirms the good quality of the fit. 

So far, the information from the signal envelope~\eqref{FitGaussianEnvelope} has not been fully exploited. From the measurements shown in Fig.~\ref{fig:AmplitudeDetuningPlot_ComparisonWithEwen} the presence of a quadratic term in the tune variation is clearly visible. However, in Eq.~\eqref{eq:amplitude-detuning} there is no such quadratic term. It can be observed that this difficulty can be overcome considering that in a case similar to that of Eq.~\eqref{Fit_AmplitudeDetuning_Tune} the parabolic tune dependence on action can be locally linearized as $\mu + \mu_2 \bar{J}_u$ at the amplitude given by $\bar{J}_u$. If multiple action values are probed, then it is possible to reconstruct the coefficients $\mu$ and $\mu_2$ using the proposed approach of fitting the signal envelope. Alternative approaches consist of determining the tune from the Hilbert normalized signal (as proposed here) or using the NAFF technique on the amplitude-modulated signal (as performed in 2012). Finally, for the uncorrected configuration, it is also possible to estimate the amplitude detuning using the LHC MAD-X model, which is based on the measured field errors of the ring magnets. In this case, the detuning coefficients are obtained from the tracking data. Note that for the corrected configuration, the MAD-X model was matched with the experimental data, which prevents it from being used for comparisons with independent beam measurement results.

\begin{figure}
    \centering
    \includegraphics[trim=42mm 30mm 51mm 42mm,width=\columnwidth,clip=]{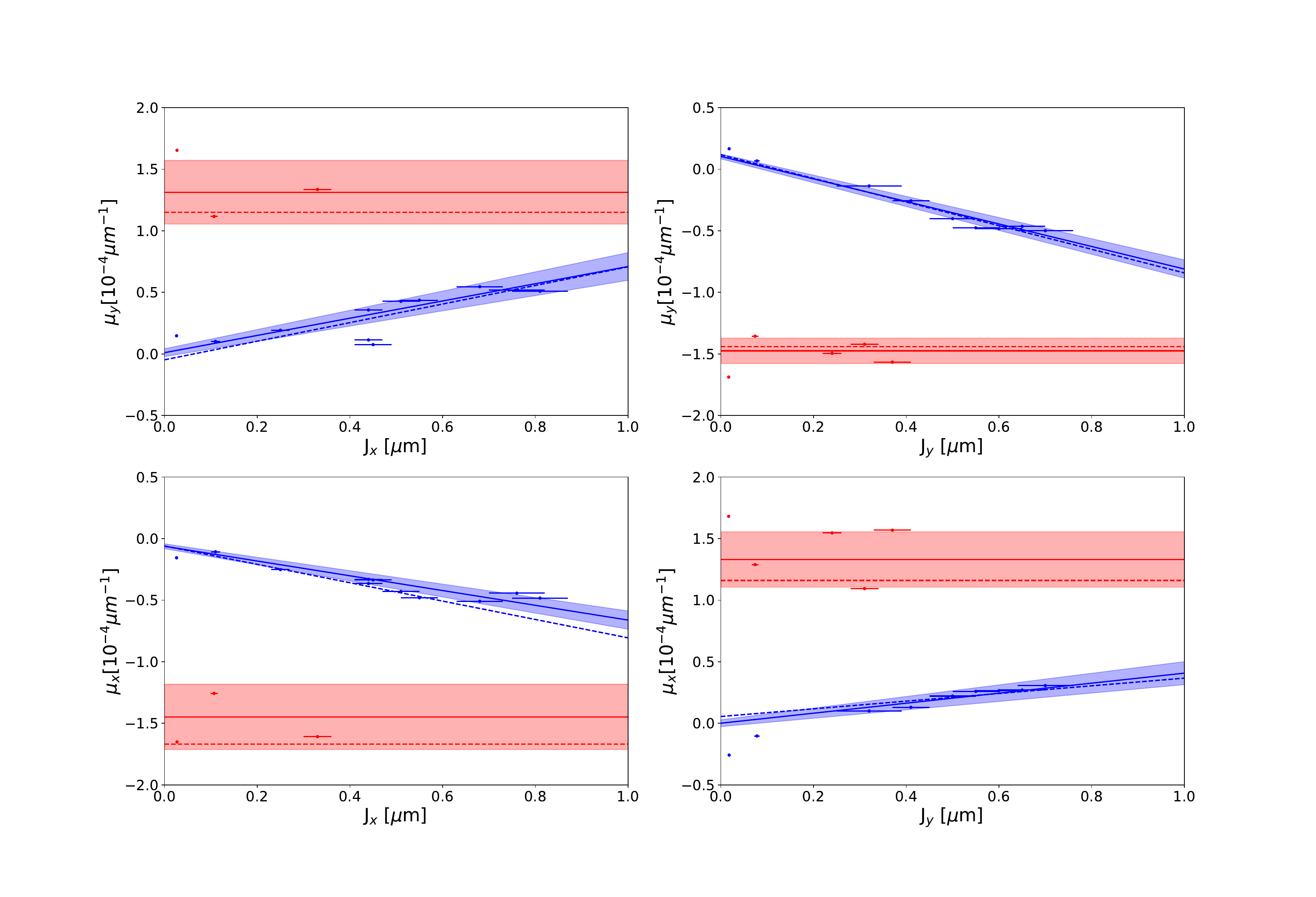}
\caption{Amplitude detuning in the horizontal (bottom row) and vertical (top row) planes is reported as a function of the horizontal (left column) and vertical (right column) single-particle action for the uncorrected (red) and corrected (black) configurations of the LHC {using the} data collected in 2012. Continuous lines refer to the fit procedure using Eq.~\eqref{Fit_AmplitudeDetuning} and the corresponding fit errors are shown with colored bars. The data points and the corresponding error bars are obtained from the proposed fit of the signal envelope. The dashed lines refer to the amplitude detuning evaluated as the derivative of the curve that fits the tune as a function of the action in Fig.~\ref{fig:AmplitudeDetuningPlot_ComparisonWithEwen}. The nice agreement between the various methods to determine the amplitude detuning is clearly visible.}
\label{fig:ReconstructionDetuningCoefficient}
\end{figure}

The results of the first three approaches are shown in Fig.~\ref{fig:ReconstructionDetuningCoefficient}, where the plotted curve is given by 
\begin{equation}\label{Fit_AmplitudeDetuning}
    \mu_u (J_u) = \mu + \mu_2 J_u \, .
\end{equation}

As shown in Fig.~ \ref{fig:ReconstructionDetuningCoefficient}, the uncorrected configuration has an amplitude detuning that does not depend on the action. The corrected configuration shows a linear trend of the horizontal and vertical amplitude detuning coefficients as a function of $J_x$ or $J_y$. We stress that the linear dependence observed of $\mu $ on the action corresponds to the measurement of a second-order amplitude detuning. Once again, there is generally good agreement between the results of the various methods applied to the experimental data. Some outliers {are} observed at very small amplitudes, {which are} explained by a weaker decoherence that makes the fit of the signal envelope inaccurate. Therefore, even second-order terms can be accurately measured by the new method based on the signal envelope.

The results of the fit procedure, including also the fit errors, are summarized in Table~\ref{Table_ResultsFit_AmplitudeDetuningCoefficient}, where also the reconstruction of the amplitude detuning from the 2012 data analysis and the LHC MAD-X model is reported. 

\begin{table*}
    \centering
    \caption{Value of $\mu$ and $\mu_2$ for the various measurement strategies, and LHC configurations, i.e., {corrected or uncorrected}. ``N.A.'' stands for ``Not Applicable'' and refers to the cases where a linear fit of the tune-action data has been performed or to the results of the MAD-X simulations that are not available for the corrected {LHC} case.}
    \label{Table_ResultsFit_AmplitudeDetuningCoefficient}
\begin{tabular}{c|c|c|c|c|c}
\toprule 
\multicolumn{2}{c|}{{$\mu$} [\SI{1e{-4}}{\micro \meter^{-1}}]} & Fitted envelope & Fitted tune vs action & Fitted tune vs action & MAD-X \\
\multicolumn{2}{c|}{}                    &                 &                       & (2012)                &       \\
\midrule 
\multirow{2}{*}{$\mu_y$ vs $J_x$}        & Uncorrected     & $1.3 \pm 0.3$         & $1.2 \pm 0.2$         & $0.8 \pm 0.1$  & $0.88 \pm 0.06$ \\
& Corrected & $-0.01 \pm 0.03$  & $-0.05 \pm 0.02$ & $-0.07 \pm 0.02$  & N.A. \\
\midrule 
\multirow{2}{*}{$\mu_y$ vs $J_y$}        & Uncorrected     & $-1.5 \pm 0.1$        & $-1.44 \pm 0.02$      & $-1.64 \pm 0.02$ & $-1.53 \pm 0.04$ \\
& Corrected & $0.10 \pm 0.02$  & $0.12 \pm 0.09$  & $0.12 \pm 0.04$   & N.A. \\
\midrule 
\multirow{2}{*}{$\mu_x$ vs $J_x$}        & Uncorrected     & $-1.4 \pm 0.3$        & $-1.5 \pm 0.3$        & $-1.16 \pm 0.03$ & $-1.08 \pm 0.03$ \\
& Corrected & $-0.06 \pm 0.02$ & $-0.01 \pm 0.02$ & $-0.04 \pm 0.05$ & N.A. \\
\midrule 
\multirow{2}{*}{$\mu_x$ vs $J_y$}        & Uncorrected     & $1.3 \pm 0.2$         & $1.16 \pm 0.07$       & $1.0 \pm 0.2$  & $0.88 \pm 0.06$  \\
&  Corrected & $0.01 \pm 0.03$ & $0.06 \pm 0.02$ & $0.08 \pm 0.03$ & N.A. \\
\midrule
\multicolumn{2}{c|}{{$\mu_2$} [\SI{1e{-2}}{\micro \meter^{-2}}]}  &  &  &  &  \\
\midrule 
\multirow{2}{*}{$\mu_y$ vs $J_x$} & Uncorrected & N.A. & N.A. & $1.4 \pm 0.4$ & $0.7 \pm 0.3$ \\
& Corrected & $0.7 \pm 0.1$ & $0.76 \pm 0.15$ & $0.30 \pm 0.05$ & N.A. \\
\midrule 
\multirow{2}{*}{$\mu_y$ vs $J_y$} & Uncorrected & N.A. & N.A. & $-0.7 \pm 0.2$ & $-0.2 \pm 0.2$ \\
& Corrected & $-0.91 \pm 0.05$ & $-0.9 \pm 0.1$ & $-0.8 \pm 0.1$ & N.A. \\
\midrule 
\multirow{2}{*}{$\mu_x$ vs $J_x$} & Uncorrected & N.A. & N.A. & $-2 \pm 1$ & $-0.6 \pm 0.1$ \\
& Corrected & $-0.60 \pm 0.05$ & $-0.8 \pm 0.1$ & $-0.9 \pm 0.3$ & N.A. \\
\midrule 
\multirow{2}{*}{$\mu_x$ vs $J_y$} & Uncorrected & N.A. & N.A. & $0 \pm 1$ & $-0.4 \pm 0.4$ \\
& Corrected & $0.41 \pm 0.07$ & $0.3 \pm 0.1$ & $0.4 \pm 0.1$ & N.A. \\
\bottomrule

\end{tabular}
\end{table*}

In inspecting the numerical values of the reconstructed physical parameters, the overall good agreement between the various approaches is clearly visible. It is also remarkable that the values obtained from the envelope analysis or those from the normalized signal are closer together than those based on the original analysis. 


\section{Conclusions} \label{sec:conc}

In this paper, we discussed the analysis of non-stationary signals with the goal of extracting useful physical information by applying harmonic analysis. This topic is particularly relevant for applications in beam dynamics and is rarely considered. In fact, most of the theoretical tools developed so far to extract accurate information about the tune of a given orbit are, strictly speaking, not applicable to the case of non-stationary signals, i.e. signals whose amplitude varies in time. 

{In this article, we proposed new approaches, based on closed-form expressions to derive the tune and the damping parameter, or on the use of the Hilbert transform. The first approach is suitable for a restricted number of physical cases, as it requires the capability of analytically computing the form of the Fourier amplitudes for a given signal form. This can be achieved for the case of pure exponential damping of a harmonic signal. Whenever closed-form expressions have been provided, the corresponding accuracy, i.e. the error on the determination of the tune and of the damping coefficient, has been determined analytically. The second approach based on the Hilbert transform is of general use and, strictly speaking, is the only correct method 
{under the condition of the Bedrosian theorem Ref.~\cite{bedrosian:1963} }

It is used to determine the signal envelope, which is then used to normalize the original signal. At this stage, high-accuracy interpolated DFT or interpolated DFT with Hanning filter methods can be applied to the normalized signal. Both methods, i.e., the analytical one for an exponential damping and that based on the Hilbert transform, have been studied in detail on sample signals to assess their performance.} 

{The appeal of the methodology based on the Hilbert transform lies in its applicability to a wide spectrum of generic non-stationary signals for the determination of their envelopes. Should the analytical representation of the envelope be ascertained, based on an understanding of the underlying physical phenomena, the envelope can subsequently be employed to determine the values of the physical parameters through a fitting procedure.}

For the specific case of amplitude detuning measurements, which is a particularly relevant application for storage rings and colliders, the proposed method allows for extracting more accurate information about the tune, and hence of the amplitude detuning, not only from a direct tune measurement as a function of the transverse amplitude but also from the decoherence properties of the signal envelope. Generally speaking, a unique beam measurement is needed to compute the amplitude detuning using the proposed method. In practice, several measurements are needed, but the possibility of extracting more information from each measurement allows the overall procedure to be improved.

The proposed method based on the Hilbert transform has been applied to {the available LHC data collected in 2012, and the results of the new analysis have been compared with the original published} results. A global agreement has been found between the results of the various techniques. 

{This promising result encourages us to promote the Hilbert transform method as a new standard, not only for the LHC but, more generally, for storage rings. This method opens up several novel applications, such as monitoring the reproducibility of nonlinear effects in a storage ring by looking at the reproducibility of the damping properties of the turn-by-turn position signal. This would be a parasitic measurement when injection oscillations are used. Furthermore, the variation of nonlinear effects along a bunch train, generated, e.g., by electron-cloud effects, could be probed by displacing the bunches in the transverse directions and measuring the damping properties of the turn-by-turn position signal.} 

\clearpage
\appendix

\section{Closed-form solution for $\nu$ and $\lambda$ for exponentially damped signals using an interpolated {DFT} approach}\label{AppendixInterpolated_FFT_DampedExponential}

Considering a damped exponential signal with frequency $\nu$ defined as:
\begin{equation}\label{DefinitionDampedExp_Appendix}
    z(n) = e^{-\lambda n} e^{2\pi i \nu n} \, .
\end{equation}
To determine a closed-form solution to determine the tune, consists in computing the {DFT} coefficients and solving and appropriate equation for the tune. First, it is possible to write the {DFT} coefficients as:
\begin{equation}\label{Phi_FFT_Coeff_DampedExp_Appendix}
    \phi (\nu _j) = \frac{1}{N} \sum_{n=1}^{\text{N}} e ^{2 \pi i (\nu - \nu _j) - \lambda n} ~,
\end{equation}
where 
\begin{equation}
    {\nu_j = \frac{j}{N}} \, . 
\end{equation}
Equation~\eqref{Phi_FFT_Coeff_DampedExp_Appendix} can be written as
\begin{equation}\label{Phi_Simplified_DampedExp_Appendix}
    \left| \phi(\nu_j) \right|^2 = \frac{1}{N^2} e^{-\lambda(N+1)} \frac{\sin^2 \left( \pi N \Delta\nu_j \mp \pi \right) + \sinh^2 \frac{\lambda N }{2}   }{\sin^2 \left( \pi \Delta\nu_j \mp \pi \right) + \sinh^2 \frac{\lambda }{2}} ~,
\end{equation}
where 
\begin{equation}
    {\Delta\nu_j = \nu - \nu _j} \, .
\end{equation}

It is possible to solve Eq.~\eqref{Phi_Simplified_DampedExp_Appendix} in terms of $\nu$ and $\lambda$ if, assuming that the maximum of $\vert \phi(\nu_j) \vert$ corresponds to 
\begin{equation}
    {\nu_k = \frac{k}{N}} \, ,
\end{equation} 
we use the expressions for $|\phi(\nu_k)|$ and $|\phi(\nu_{k \pm 1})|$. It is then possible to show that 
\begin{equation}\label{Tune_DampedExponential_Appendix}
    \begin{split}
       \nu_{\pm} & = \frac{k}{N}  + \frac{1}{\pi} \arctan \biggl[  \frac{1}{\tan \frac{\pi}{N}} \biggl( 
   \frac{\eta +1}{\eta -1} \pm \\
   &\sqrt{ \left( \frac{\eta +1}{\eta -1}  \right)^2 + \tan ^2  \frac{\pi}{N} } \biggr) \biggr] ~,
    \end{split}
\end{equation}
where
\begin{equation}
    \eta = \frac{\chi_+ -1}{\chi_- -1} ~, \hspace{1cm} \chi_\pm = \frac{|\phi(\nu_k)|^2}{|\phi(\nu_{k\pm1})|^2}
\end{equation}
is an estimate of $\nu$. The choice between $\nu_-$ and $\nu_+$ is based on the values of $|\phi(\nu_{k \pm 1})|$: If $|\phi(\nu_{k - 1})| > |\phi(\nu_{k + 1})|$ then $\nu_-$ should be selected; otherwise, $\nu_+$ should be selected. 

Similarly, it is possible to show that the following holds
\begin{equation}\label{LambdaReconstruction_DampedExp_Appendix}
\begin{split}
     \lambda_{\pm} & = 2 \arcsinh \\
     & \sqrt{ \frac{|\phi(\nu_{k \pm 1})|^2 \sin^2 \left( \pi \Delta\nu_k \mp \frac{\pi}{N} \right) -  |\phi(\nu_{k})|^2 \sin^2 \pi \Delta\nu_k }       {|\phi(\nu_{k})|^2 - |\phi(\nu_{k \pm 1})|^2}} \, ,
\end{split}
\end{equation}
where 
\begin{equation}
    {\Delta\nu_k = \nu - \nu _k} \, , 
\end{equation}
the choice between the two values of $\lambda$ is made in a similar way as for the tune (additional details of the determination of Eqs.~\eqref{Tune_DampedExponential_Appendix} and~\eqref{LambdaReconstruction_DampedExp_Appendix} can be found in Ref.~\cite{Fabre:2043807}). 

The determination of the precision with which Eqs. \eqref{Tune_DampedExponential_Appendix} and~\eqref{LambdaReconstruction_DampedExp_Appendix} provide the tune and the damping factor is carried out by perturbing the {DFT} coefficients due to the inclusion of an additional frequency (as done in Ref.~\cite{Bartolini:292773}). The perturbed {DFT} coefficient will read
\begin{equation}\label{phi_perturbed_DampedExp_AppendixA}
    \hat{\phi}(\nu_j) = \phi(\nu_j) + O\left ( \frac{1}{N} \right ) \, ,
\end{equation}
and it is immediate to obtain
\begin{equation}
\begin{split}
    \vert \hat{\phi}(\nu_j) \vert & = \vert \phi(\nu_j) \vert +  O\left ( \frac{1}{N} \right ) \\
    \vert \hat{\phi}(\nu_j) \vert^2 & = \vert \phi(\nu_j) \vert^2 +  O\left ( \frac{1}{N^2} \right ) \, ,    
\end{split}
\end{equation}
where the second relation is obtained by noting that $\phi(\nu_j) \approx O\left ( \frac{1}{N} \right )$.

Using the previous relationships and replacing them in Eqs.~\eqref{Tune_DampedExponential_Appendix} and~\eqref{LambdaReconstruction_DampedExp_Appendix} one obtains the following results
\begin{equation}
    \begin{split}
        \hat{\nu} & = \nu + O \left ( \frac{1}{N^2}\right ) \\
        \hat{\lambda} & = \lambda  + O \left ( \frac{1}{N^2}\right ) \, ,
    \end{split}
\end{equation}
which provide an estimate of the error affecting the closed-form expressions for $\nu$ and $\lambda$ as a function of $N$.

\section{Closed-form solution for $\nu$ and $\lambda$ for exponentially damped signals using an interpolated {DFT} with Hanning filter approach}\label{Appendix_HanningDampedExponential}

For the case of a signal including the Hanning filter, one obtains that the {DFT} coefficients read
\begin{equation} \label{AppE_phiNC_Hann}
    \phi (\nu_j) = \frac{2}{N} \sum_{n=1}^N e^{2 \pi i (\nu - \nu_j) n - \lambda n} \sin^2\frac{\pi n}{N} \, ,
\end{equation}
which can be cast in the form
\begin{equation}\label{AppE_phi_Hanning_ComplexNotation}
    \phi (\nu_j) = \frac{i e^{-\lambda N} \sin ^2\frac{\pi }{N} \cot~\theta_j(\lambda)  \left(-e^{\lambda  N}+e^{2 \pi i  (\nu - \nu_j) N}\right)} {N \left(\cos \frac{2 \pi }{N}-\cos 2\theta_j(\lambda)\right)} \, ,
\end{equation}
where $\theta_j(\lambda)$ is a complex variable defined as
\begin{equation}\label{ThetaDefinition_AppE}
     \theta_j(\lambda) = \pi (\nu - \nu_j) +i \frac{\lambda}{2 } \, .
\end{equation}

It is possible to find a closed-form solution of Eq.~\eqref{AppE_phi_Hanning_ComplexNotation} for $\nu$ and $\lambda$ considering Eq.~\eqref{AppE_phi_Hanning_ComplexNotation} for $\nu_k = k/N$, representing the maximum of the {DFT} spectrum and its neighboring coefficients for $k\pm 1$. 

In fact, it is possible to obtain two trigonometric equations in the complex domain in the unknown $\theta_k(\lambda)$ that read
\begin{equation}\label{CosineSineEquation1}
\begin{split}
    \left(\cos \frac{2\pi}{N} - \chi_+ \right) \sin 2\theta_k(\lambda) - & \sin \frac{2\pi}{N} \cos 2\theta_k(\lambda) \\
    - & \sin\frac{2\pi}{N} (\chi_++1) =0 ~,    
\end{split}
\end{equation}
or 
\begin{equation}\label{CosineSineEquation2}
\begin{split}
    \left(\cos \frac{2\pi}{N} - \chi_- \right) \sin 2\theta_k(\lambda) + & \sin \frac{2\pi}{N} \cos 2\theta_k(\lambda) \\
    + & \sin\frac{2\pi}{N} (\chi_-+1) =0 \, , 
\end{split}
\end{equation}
where $\chi_\pm$ is defined as
\begin{equation}\label{Chi_SimplifiedForm}
     \chi_\pm = \frac{\phi(\nu_k)}{\phi(\nu_{k \pm 1})} \, .
\end{equation}

The two equations can be summed up to obtain
\begin{equation}
\begin{split}
    \left [2 \cos \frac{2\pi}{N} - \left (\chi_+ + \chi_- \right )\right ] \sin 2\theta_k(\lambda) = \sin\frac{2\pi}{N} (\chi_+ - \chi_-) \, ,
\end{split}
\end{equation}
from which one obtains 
\begin{equation}
    \theta_k(\lambda) = \frac{1}{2} \arcsin \left [\frac{\sin\frac{2\pi}{N} (\chi_+ - \chi_-)}{2 \cos \frac{2\pi}{N} - \left (\chi_+ + \chi_- \right )} \right ] \, . 
\end{equation}

To determine the tune and damping factor, it is necessary to determine the inverse sine function of a complex number. It is possible to find its definition in Ref.~\cite{abramowitz} using the formula 
\begin{equation}\label{AbramowitzFormula}
\begin{split}
     \arcsin(z)&  = \arcsin(x + i y) = k \pi + (-1)^k \arcsin \beta + \\
     & + (-1)^k i \ln \left( \alpha + \sqrt{(\alpha^2 + 1)}\right) \quad k \in \mathbb{N}_0
\end{split}
\end{equation}
and
\begin{equation}
\begin{split}
    \alpha & = \frac{1}{2} \sqrt{(x+1)^2 + y^2} + \frac{1}{2} \sqrt{(x-1)^2 + y^2} \\
    \beta & = \frac{1}{2} \sqrt{(x+1)^2 + y^2}- \frac{1}{2} \sqrt{(x-1)^2 + y^2} \, .    
\end{split}
\label{eq:alpha-beta}
\end{equation}

Thus, from the following equalities
\begin{equation}
    2\pi(\nu - \nu_k) + i \lambda =  \arcsin \beta + i \ln \left( \alpha + \sqrt{\alpha^2 + 1}\right)
\end{equation}
one obtains
\begin{equation}\label{TuneEq_Hanning_AppE}
\begin{split}
    \nu & = \frac{k}{N} + \frac{1}{2\pi} \arcsin \beta \\
    \lambda & = \ln \left( \alpha + \sqrt{\alpha^2 + 1}\right) \, .
\end{split}
\end{equation}

Also in this case, it is possible to determine the error scaling law by determining $\nu$ and $\lambda$. The approach is similar to what is used for the case of interpolated {DFT} of an exponentially damped signal. Adding a frequency to the original signal for which the closed-form solutions have been determined implies that the perturbed {DFT} coefficient reads
\begin{equation}\label{phi_perturbed_DampedExp_AppendixB}
    \hat{\phi}(\nu_j) = \phi(\nu_j) + O\left ( \frac{1}{N^3} \right ) \, ,
\end{equation}
and it is immediate to obtain
\begin{equation}
    \hat{\chi}_{\pm} = \chi_{\pm} +  O\left ( \frac{1}{N^3} \right ) \, ,
\end{equation}
and similarly, one finds 
\begin{equation}
    {\hat{z} = z + O \left (\frac{1}{N^3} \right )} \, , 
\end{equation}
and finally, using Eqs.~\eqref{eq:alpha-beta} and~\eqref{TuneEq_Hanning_AppE} one obtains the final estimate
\begin{equation}
    \begin{split}
        \hat{\nu} & = \nu + O \left ( \frac{1}{N^3}\right ) \\
        \hat{\lambda} & = \lambda  + O \left ( \frac{1}{N^3}\right ) \, ,
    \end{split}    
\end{equation}
\bibliographystyle{unsrt}
\bibliography{TuneVaryingAmplitude.bib}

\end{document}